\documentclass[useAMS,usenatbib]{mn2e}
\usepackage{epsfig}
\usepackage{times}

\newcommand{\RS}{\rho_{_{\mathrm{S}}}}
\newcommand{\RL}{\rho_{_{\mathrm{L}}}}
\newcommand{\LD}{S(\frac{r}{\RS})}
\newcommand{\LLD}{S\left(\frac{r}{\RS}\right)}
\newcommand{\AERR}{\theta_{\mathrm{E}}}
\newcommand{\AS}{\theta_{{\mathrm{S}}}}
\newcommand{\AL}{\theta_{{\mathrm{L}}}}
\newcommand{\LDC}{\Gamma_{\mathrm{S}}}
\newcommand{\fls}{f_{_{\mathrm{LS}}}}
\newcommand{\Dol}{D_{_{\mathrm{OL}}}}
\newcommand{\Dos}{D_{_{\mathrm{OS}}}}
\newcommand{\ML}{M_{_{\mathrm{L}}}}
\usepackage{natbib}
\usepackage{color}

\begin{document}

\title[Astrometric microlensing]{Finite-source and finite-lens effects in astrometric microlensing}
\author[C.-H. Lee, S. Seitz, A. Riffeser and R. Bender]{C.-H. Lee$^{1}$\thanks{E-mail:chlee@usm.lmu.de}, S. Seitz$^{1,2}$, A. Riffeser$^{1}$ and R. Bender$^{1,2}$\\
$^{1}$University Observatory Munich, Scheinerstrasse 1, 81679 M\"unchen, Germany\\
$^{2}$Max Planck Institute for Extraterrestrial Physics, Giessenbachstrasse, 85748 Garching, Germany}

\date{Accepted --. Received --; in original form 2010 January 14}

\pagerange{\pageref{firstpage}--\pageref{lastpage}} \pubyear{0000}

\maketitle

\label{firstpate}

\begin{abstract}
The aim of this paper is to study the astrometric trajectory of microlensing events with an 
extended lens and/or source. We consider not only a dark lens but also a luminous lens as well. 
We find that the discontinuous finite-lens trajectories given by Takahashi will 
become continuous in the finite-source regime.
The point lens (source) approximation alone gives an under (over)estimation of the 
astrometric signal when the size of the lens and source are not negligible. While the finiteness of the 
source is revealed when the lens transits the surface of the source,
the finite-lens signal is most prominent when the lens is 
very close to the source. Astrometric microlensing towards the Galactic bulge, 
Small Magellanic Cloud and M31 are discussed, which indicate that the finite-lens effect is beyond the detection limit 
of current instruments. Nevertheless, it is possible to distinguish between 
self-lensing and halo lensing through a (non-)detection of the astrometric ellipse. We also consider the 
case where the lens is luminous itself, as has been observed where a lensing event was followed up with the
\textit{Hubble Space Telescope}. We show that the astrometric signal will be reduced in a 
luminous-lens scenario. The physical properties of the event, such as the lens-source flux ratio, 
the size of the lens and source nevertheless can be derived by fitting the astrometric trajectory.
\end{abstract}


\section{Introduction}

Most of the microlensing events detected to date are through photometric monitoring of the source 
flux. In this case, the information on the physical identity of the lens is reduced, because the only 
quantity one can retrieve from the light curve is the Einstein timescale $t_{\mathrm{E}}$. $t_{\mathrm{E}}$ 
is defined by the time required for the source to transit the angular Einstein radius 
$\AERR$ of the lens \citep{2000ApJ...542..785G}:
\begin{equation}
    t_{\mathrm{E}} = \frac{\AERR}{|\mbox{\boldmath $\mu_{rel}$}|},\quad 
    \AERR = \sqrt{k\ML\pi_{rel}},\quad
    k\equiv \frac{4G}{c^2 \mathrm{AU}}\approx 8.14\frac{\mathrm{mas}}{M_{\odot}}, 
    \label{eq.tE}
\end{equation}
where \mbox{\boldmath $\mu_{rel}$} is the relative lens-source proper motion, $\ML$ is the mass of the 
lens, $\pi_{rel}:= \mathrm{AU}/(\Dol^{-1}-\Dos^{-1})$ is the relative lens-source parallax,
$\Dol$ and $\Dos$ are distance to the lens and the source from the observer, respectively. 
Equation~(\ref{eq.tE}) shows that the mass, distance, and velocity of the lens are degenerated 
into $t_{\mathrm{E}}$. 

To better constrain the lens properties, \cite{1995A&A...294..287H}, 
\cite{1995ApJ...453...37W} and \cite{1995AJ....110.1427M} thus suggested to use astrometric 
microlensing. That is, to measure the centroid displacement of the two images during the course 
of microlensing. Former studies have shown that the trajectory of the centroid displacement will 
trace out an ellipse, and the size of the ellipse is proportional to the angular Einstein radius. 
Therefore, one can determine $\AERR$ through the observation of such astrometric 
ellipses and constrain the relative lens-source proper motion. 
\cite{1992ApJ...392..442G} has shown that if one can further measure the microlens parallax 
$\pi_{\mathrm{E}}=\sqrt{\pi_{rel}/(k\ML)}$ form the light-curve distortion induced by the 
orbital motion of the Earth, the lens mass $\ML$ and the relative lens-source parallax $\pi_{rel}$ can 
be determined without ambiguity: 
\begin{equation}
  \ML = \frac{\AERR}{k\pi_{\mathrm{E}}},\quad
  \pi_{rel}=\pi_{\mathrm{E}}\AERR.
  \label{eq.M}
\end{equation}

The location of the lens can be derived as well if the distance to the source is well known, 
which is often the case towards the Galactic bulge and Magellanic Clouds.

The typical value of the astrometric microlensing signal for a source in the Galactic bulge 
and a 0.5 $M_{\odot}$ lens located half-way to the source is of order of 0.1 mas, which is much larger 
than the astrometric accuracy of upcoming space missions such as   
\textit{Space Interferometry Mission} \citep[\textit{SIM}; ][]{1997SPIE.2871..504A}, 
\textit{Global Astrometric Interferometer for Astrophysics} \citep[\textit{GAIA}; ][]{1994SPIE.2200..599L} and 
ground-based instruments, e.g. \textit{Phase Referenced Imaging and Micro-arcsecond Astrometry} 
\citep[\textit{PRIMA}; ][]{1998SPIE.3350..807Q}. 
\textit{GAIA} will survey the whole sky with sources brighter than 20 mag in \textit{V} band. 
It is expected to reach an astrometric accuracy of 30 $\mu$as (150 $\mu$as) with $V <$ 12 ($V <$ 16) 
for a single measurement \citep{2002MNRAS.331..649B} and an estimated detection of $\approx$ 1000 
events \citep{2000ApJ...534..213D}. Unlike \textit{GAIA}, which only scans the sky with a pre-determined 
pattern, \textit{SIM} can point to selectable targets and thus tracks the ongoing microlensing 
event upon request. The expected accuracy of \textit{SIM} is 5 $\mu$as (20 $\mu$as) for $V <$ 12 
($V < $ 16) with 1-hour integration time \citep{2008SPIE.7013E.151G}. While \textit{SIM} and 
\textit{GAIA} are scheduled to launch in the next few years, \textit{PRIMA} has already been 
installed on the Very Large Telescope Interferometer (VLTI) and aims at achieving 10-$\mu$as accuracy level in 30 min 
integration time provided a reference star within 10 arcsec and a 200-m baseline 
\citep{2008NewAR..52..199D}.

In addition to the standard point-source point-lens (PSPL) microlensing, single-lens events revealing 
an extended source signal have also been observed photometrically 
\citep[e.g.][]{1997ApJ...491..436A, 2004ApJ...617.1307J, 2004ApJ...603..139Y, 2006A&A...460..277C, 2009A&A...508..467B, 2009ApJ...703.2082Y, 2009arXiv0912.2312Z, 2010A&A...518A..51F}. 
\cite{1998MNRAS.300.1041M} thus derived the astrometric trajectory of finite-source events with a 
point-lens (FSPL). On the other hand, \cite{2003ApJ...595..418T} studied the centroid displacement 
of finite-lens effects but assuming a point-source (PSFL). Furthermore, \cite{2002ApJ...579..430A} 
and \cite{2009ApJ...695..200L} have  
investigated the combination of finite-source and finite-lens (FSFL) effects photometrically, but left 
aside the astrometric aspect. Since the FSFL light curve deviates from either the PSFL or FSPL, as shown by 
\cite{2002ApJ...579..430A}, we are motivated to study the astrometric behaviour when both FS and 
FL effects are relevant. There are events where both the source and the lens are resolved by 
the \textit{Hubble Space Telescope (HST)}. This implies that the lens can also be a star and implies a luminous-lens 
scenario rather than a dark lens \citep[][]{2001Natur.414..617A, 2007ApJ...671..420K}, which is also the case for 
self-lensing. We thus consider the light contribution from the lens star and study the astrometric behaviour 
by allowing for a luminous lens. 

This paper is organized as follows. In \textsection~2 we introduce the theory of astrometric 
microlensing. We take into account the FS effects either with a uniform surface brightness 
source or with a more general surface brightness profile in \textsection~3. We further include a 
dark lens with finite size in \textsection~4. One might expect not only shadowing but also light 
contribution from the lens as well. Therefore, we allow for a luminous lens in \textsection~5.
The aforementioned properties of the microlensing system can be estimated by fitting 
the formula in \textsection~5.  
A discussion of possible events with sources located in the Galactic bulge, Small Magellanic Cloud (SMC) 
and M31 is presented in \textsection~6 followed by a summary in \textsection~7. 
\\

\section{Astrometric trajectory of the lensed images}

Let \mbox{\boldmath $\varphi$}$_{_\mathrm{S}}$ and \mbox{\boldmath $\varphi$}$_{_\mathrm{L}}$ be the angular position of 
the source and lens. Then one can derive the position (\mbox{\boldmath $\theta$}) and 
the amplification ($A$) of the two lensed images in the lens plane 
through the dimensionless impact parameter \mbox{\boldmath $u$} := (\mbox{\boldmath $\varphi$}$_{_\mathrm{S}}$ - 
\mbox{\boldmath $\varphi$}$_{_\mathrm{L}}$)/$\AERR$ \citep[see e.g. ][]{1993A&A...278L..27H, 
1995A&A...294..287H, 1995ApJ...453...37W, 1995AJ....110.1427M}:

\begin{equation}
    \mbox{\boldmath $\theta$}_{\pm} = \frac{1}{2}\left[u\pm\sqrt{u^2+4}\right]\hat{\mbox{\boldmath $u$}},\quad A_{\pm} = \frac{1}{2}\left[\frac{u^2+2}{u\sqrt{u^2+4}} \pm 1\right],
\end{equation}

where $u$ = $|\mbox{\boldmath $u$}|$ and $\hat{\mbox{\boldmath $u$}}$ = $\mbox{\boldmath $u$}/u$. 
Note that $\mbox{\boldmath $\theta$}_\pm$ and $\hat{\mbox{\boldmath $u$}}$ are vectors 
while $A_\pm$ are scalars. The centroid of the images can be calculated by weighting the position 
of the two images with their amplification:

\begin{equation}
  \mbox{\boldmath $\theta$}_{c,\mathrm{PSPL}} = \frac{A_+ \mbox{\boldmath $\theta_+$} + A_- \mbox{\boldmath $\theta_-$}}{A_++A_-}=\frac{1}{2}\left[\frac{u(u^2+4)}{u^2+2}+u\right]\hat{\mbox{\boldmath $u$}}
\end{equation}

and the centroidal shift relative to the source is

\begin{equation}
    \delta \mbox{\boldmath $\theta$}_{c,\mathrm{PSPL}} = \mbox{\boldmath $\theta_c - u$} = \frac{u}{u^2+2}\hat{\mbox{\boldmath $u$}}.
    \label{eq.thetaC}
\end{equation}

If we neglect the parallax effects, the relative motion between the lens and source can be 
approximated by rectilinear motion so that

\begin{equation}
    u(t) = \sqrt{\tau^2 + u_0^2},\quad \tau = \frac{t-t_0}{t_E},
\end{equation}
where $u_0$ is the closest approach at $t_0$. 

\begin{figure}
  \centering
  \includegraphics[scale=0.65]{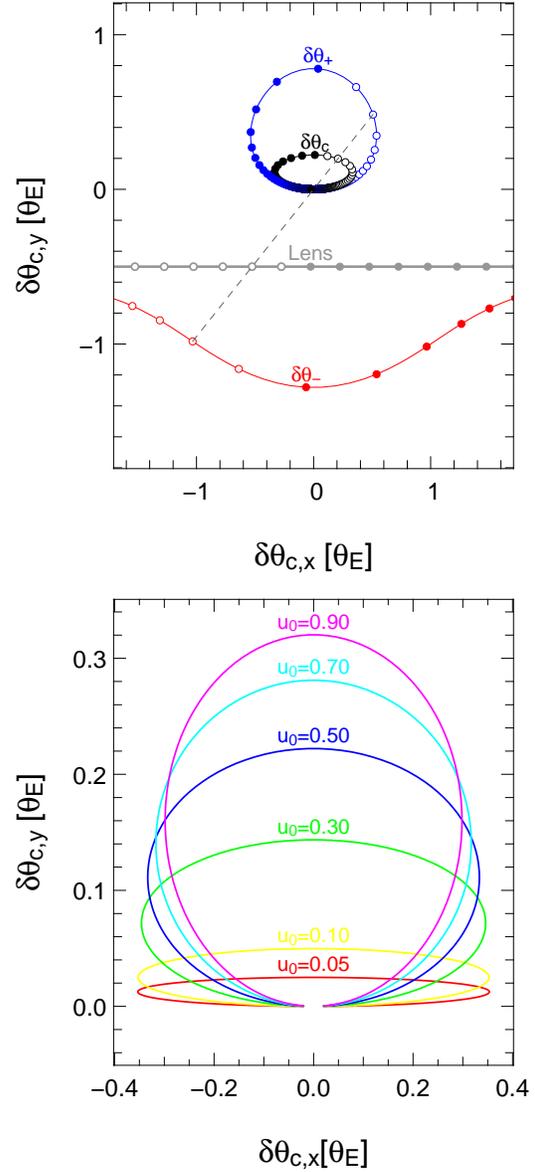}
  \caption{Centroid shifts for PSPL. Upper panel: the trajectory of the plus-image (in blue), minus-image 
           (in red), centroid of these two images (in black), and the lens 
           (in gray) relative to the source center assuming $t_0$ = 0, $t_{\mathrm{E}}$ 
           = 10 d, and $u_0$ = 0.5 $\AERR$. Lower panel: centroid displacement 
           for different values of $u_0$.}
  \label{fig.u0}
\end{figure}

The centroidal shift can then be decomposed into components parallel to 
\mbox{\boldmath $\mu_{rel}$}, $\delta\theta_{c,\mathrm{PSPL},x}$, and perpendicular to 
\mbox{\boldmath $\mu_{rel}$}, $\delta\theta_{c,\mathrm{PSPL},y}$ (see Fig.~\ref{fig.u0}). One further 
finds that the centroidal shift actually traces out an ellipse \citep{1995ApJ...453...37W} 

\begin{equation}
\left(\frac{\delta\theta_{c,\mathrm{PSPL},x}}{a}\right)^2 + \left(\frac{\delta\theta_{c,\mathrm{PSPL},y}-b}{b}\right)^2 = 1,
\end{equation}

where $a$ and $b$ are the semi-major and semi-minor axis of the ellipse, respectively,

\begin{equation}
  \begin{array}{ll}
    a = \frac{1}{2}\frac{1}{\sqrt{u_0^2+2}}, & b = \frac{1}{2}\frac{u_0}{u_0^2+2}.
  \end{array}
\end{equation}

The trajectory of centroidal shift with different values of $u_0$ is shown in Fig.~\ref{fig.u0}.

\begin{figure}
  \centering
  \includegraphics[scale=0.6]{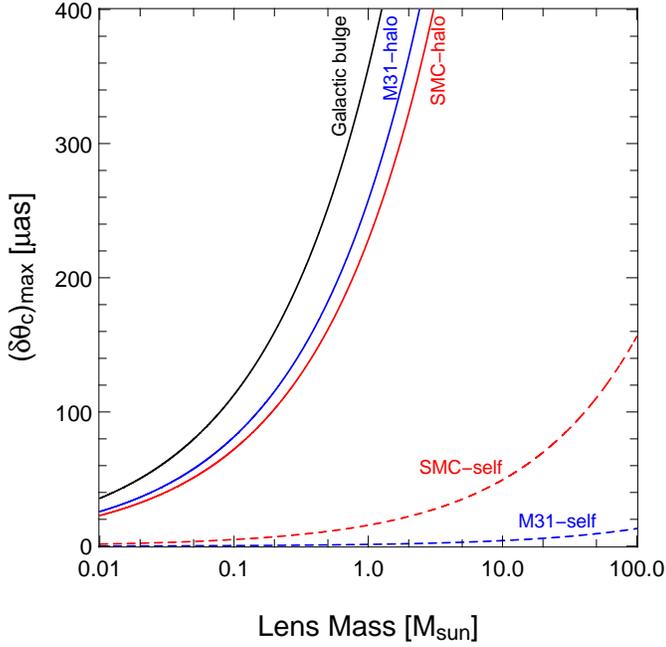}
  \caption{Maximum values for the centroidal shift versus lens mass. For illustration, we set the lens 
           of halo lensing towards SMC ($\Dos$ = 65 kpc) and M31 ($\Dos$ = 770 kpc) to be 15 kpc from 
           the observer, and 1 kpc in front of the source as self-lensing. We only show the case of 
           halo lensing for Galactic bulge ($\Dos$ = 8 kpc) assuming the lens is half-way 
           to the background source.}
  \label{fig.M_vs_thetaC}
\end{figure}

Taking the derivative of equation~(\ref{eq.thetaC}), one finds that the maximum centroidal shift occurs when 
$u = \sqrt{2}$ and has an absolute value of $\sqrt{2}/(2+2) = 1/(2\sqrt{2}) \approx 0.3536$, 
i.e. about one-third of the angular Einstein radius. For a source located in the Galactic 
bulge with a lens of 0.5 $M_{\odot}$ located half-way to the source, the angular Einstein radius 
is 712 $\mu as$, which is 1 (2) mag larger than the planned astrometric accuracy of 
the \textit{GAIA} (\textit{SIM}) mission even after taking one-third of its value. The maximum 
values for the centroidal shifts with halo and self-lensing towards SMC and M31 are shown in 
Fig.~\ref{fig.M_vs_thetaC}.

\section{The Finite-Source Effects}

For an extended source, the centroid of the two images is obtained by a two-dimensional integral 
of the image position weighted by its amplification over the surface of the source 
\citep{1995ApJ...453...37W, 1998MNRAS.300.1041M}:

\begin{equation}
\delta \mbox{\boldmath $\theta$}_{c,\mathrm{FSPL}} = \frac{\int_0^{2\pi} \int_0^{\RS} \left[A_+ \mbox{\boldmath $\theta$}_+ + A_- \mbox{\boldmath $\theta$}_- \right]\LLD rdrd\phi }{\int_0^{2\pi} \int_0^{\RS} \left[A_+ + A_- \right] \LLD rdrd\phi} - \mbox{\boldmath $u$}
\label{eq.FSPL}
\end{equation}

where $\LD$ is the source surface-brightness profile, $\RS$ := $\AS / \AERR$ 
is the source radius in units of the angular Einstein ring radius and $r$ is the distance to the 
source centre.\\

\begin{figure}
  \centering
  \includegraphics[scale=0.35]{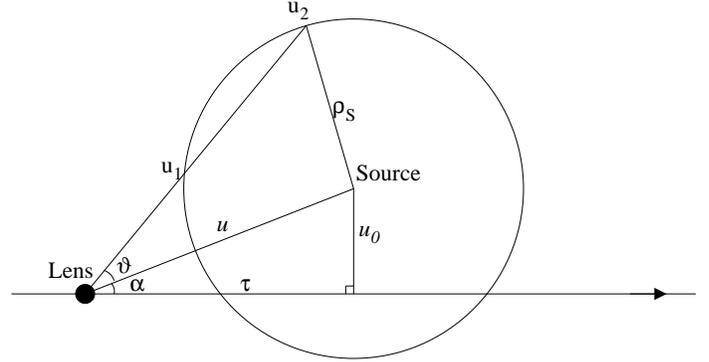}
  \caption{Schematic view of the lens-centered coordinates used in equation~(\ref{eq.FSPL1D}).}
  \label{fig.lens-source}
\end{figure}

For a source with uniform surface brightness 
[i.e. $\LD$ is constant all over the surface of the source], the integration over 
the source surface can be reduced into a 
one-dimensional integral following the lens-centered coordinates approach of \cite{2009ApJ...695..200L}.
One thus derives the values for $\delta\theta_{c,\mathrm{FSPL},x}$ and 
$\delta\theta_{c,\mathrm{FSPL},y}$

\begin{equation}
  \begin{array}{l}
    \delta \theta_{c,\mathrm{FSPL},x} = \frac{\int_0^{2\pi}\left[\frac{1}{3} \left(\tilde{u}^2+1 \right)\sqrt{\tilde{u}^2+4}\right]_{\tilde{u}=u_1}^{^{u_2}} \cos(\vartheta+\alpha)d\vartheta}{\int_0^{2 \pi} \left[ \frac{1}{2} \tilde{u}\sqrt{\tilde{u}^2+4} \right]_{\tilde{u}=u_1}^{^{u_2}}d\vartheta} - \tau, \\
    \delta \theta_{c,\mathrm{FSPL},y} = \frac{\int_0^{2\pi}\left[\frac{1}{3} \left(\tilde{u}^2+1 \right)\sqrt{\tilde{u}^2+4}\right]_{\tilde{u}=u_1}^{^{u_2}} \sin(\vartheta+\alpha)d\vartheta}{\int_0^{2 \pi} \left[ \frac{1}{2} \tilde{u}\sqrt{\tilde{u}^2+4} \right]_{\tilde{u}=u_1}^{^{u_2}}d\vartheta} - u_0, \\
  \end{array}
  \label{eq.FSPL1D}
\end{equation}

where the integration boundaries $u_1$ and $u_2$ are
\begin{equation}
  \begin{array}{l}
    u_1 =
    \left\{
      \begin{array}{ll}
        0
        &
        u\le\RS
        \\
        u \cos{\vartheta} - \sqrt{\RS^2 - u^2 \sin^2\vartheta}
        &
        u>\RS \wedge \vartheta\le \sin^{-1}(\frac{\RS}{u})
        \\
        0
        &
        u>\RS \wedge \vartheta> \sin^{-1}(\frac{\RS}{u})
      \end{array}
    \right.
    \\
    u_2 =
    \left\{
      \begin{array}{ll}
        u \cos{\vartheta} + \sqrt{\RS^2 - u^2 \sin^2\vartheta}
        &
        u\le\RS
        \\
        u \cos{\vartheta} + \sqrt{\RS^2 - u^2 \sin^2\vartheta}
        &
        u>\RS \wedge \vartheta\le \sin^{-1}(\frac{\RS}{u})
        \\
        0
        &
        u>\RS \wedge \vartheta > \sin^{-1}(\frac{\RS}{u})
      \end{array}
    \right.
  \end{array}
  \label{eq.u2}
\end{equation}

and $\alpha = \tan^{-1}(u_0 / \tau)$. The relative lens-source configuration 
and the parameters used in equations~(\ref{eq.FSPL1D})-(\ref{eq.u2}) are sketched in 
Fig.~\ref{fig.lens-source}.

\begin{figure}
  \includegraphics[scale=0.83]{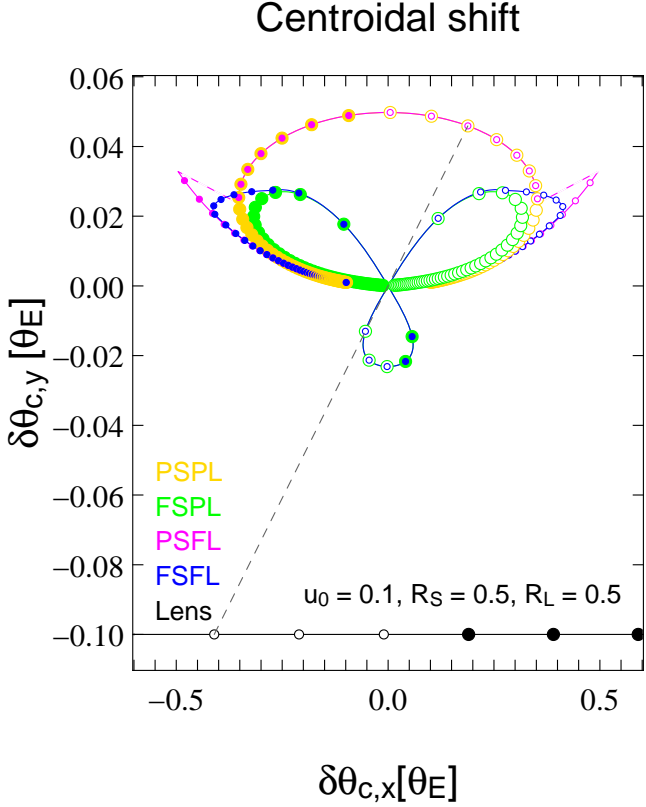}
  \caption{An example for centroidal shifts of a microlensing event 
           assuming $u_0$ = 0.1 $\AERR$, source radius = 0.5 $\AERR$ and lens radius 
           = 0.5 $\AERR$. We show the trajectory of a PSPL (in yellow), 
           FSPL (in green), PSFL (in red), and FSFL (in blue). 
           Note that the dashed red line indicates the discontinuous part of the 
           trajectory in a PSFL event.}
   \label{fig.AM}
\end{figure}

An example for an FS centroidal shift is shown in Fig.~\ref{fig.AM} along with 
the light curve in Fig.~\ref{fig.compare_FSFL}. 
The FS effect drives the centroidal shift towards the source center for small $u$, 
and the trajectory becomes cloverleaf-like when $u_0$ is smaller than the source radius 
(see Fig.~\ref{fig.FSFL}).\\ 


We also show the centroidal shifts of a limb-darkened source with an one parameter 
linear limb-darkening profile \citep{2004ApJ...603..139Y}, that is

\begin{equation}
  \LLD = \bar{S}\left[1-\LDC \left(1-\frac{3}{2}\sqrt{1-\left(\frac{r}{\RS}\right)^2} \right) \right],
  \label{eq.LD}
\end{equation}

where $r$ is the distance to the source center. $\LDC$ is the 
wavelength-dependent limb-darkening coefficient. $\bar{S}$ is the mean surface brightness of the 
source. When $\LDC$ = 0, equation~(\ref{eq.LD}) gives us a source with uniform brightness. 
The trajectory of the centroidal shift by the limb-darkened sources shows only small 
difference from that of the uniform brightness source, as shown in Figs~\ref{fig.compare_LD} and 
\ref{fig.FSFL}.

\section{The Finite-Lens Effects}

For simplicity, we begin with the case of PSFL. 
The light from the plus-image will be obscured 
by the lens if its distance to the lens is smaller than the lens radius. 
That is, the plus-image vanishes when $\theta_+ = |\mbox {\boldmath $\theta$}_+| < \RL$ 
($\RL:= \AL / \AERR$). Similarly, the minus-image vanishes 
when $\theta_- = |\mbox {\boldmath $\theta$}_-| < \RL$. Therefore, the centroidal shift 
taking into account the lens size is \citep{2003ApJ...595..418T}

\begin{equation}
  \delta \mbox{\boldmath $\theta$}_{c,\mathrm{PSFL}} = \frac{A_+ \mbox{\boldmath $\theta$}_+ \Theta(\theta_+ - \RL) + A_- \mbox{\boldmath $\theta$}_- \Theta(\theta_- - \RL)}{A_+\Theta(\theta_+ - \RL) + A_-\Theta(\theta_- - \RL)} - \mbox{\boldmath $u$},
\label{eq.PSFL}
\end{equation}

where $\Theta(x)$ is the Heaviside step function. An example for a PSFL centroidal 
shift is shown in Fig.~\ref{fig.AM} along with the light curve in Fig.~\ref{fig.compare_FSFL}. 
The trajectory is composed of two 
discontinuous parts: it follows the plus-image trajectory at larger $u$ and returns to the 
PSPL centroidal trajectory at smaller $u$. This can be explained as follows. 
When the FL effects set in, the lens first obscures the minus-image because it is always 
inside the Einstein ring and has smaller distance to the lens compared to the plus-image (which 
is always outside the Einstein ring). In addition, the value of $\theta_-$ becomes larger for 
smaller $u$, as we can see from Fig.~\ref{fig.u0}, which brings the trajectory back to the 
PSPL centroidal trajectory at smaller $u$ for smaller $\RL$. As a 
consequence, when the size of the lens increases, the trajectory tends to be more plus-image-like 
until the lens size becomes so large that it completely obscures the light even from the plus-image.\\

Combining equations~(\ref{eq.FSPL}) and (\ref{eq.PSFL}), we are able to fully consider the FSFL  
effects:
\begin{figure}
  \includegraphics[scale=0.66]{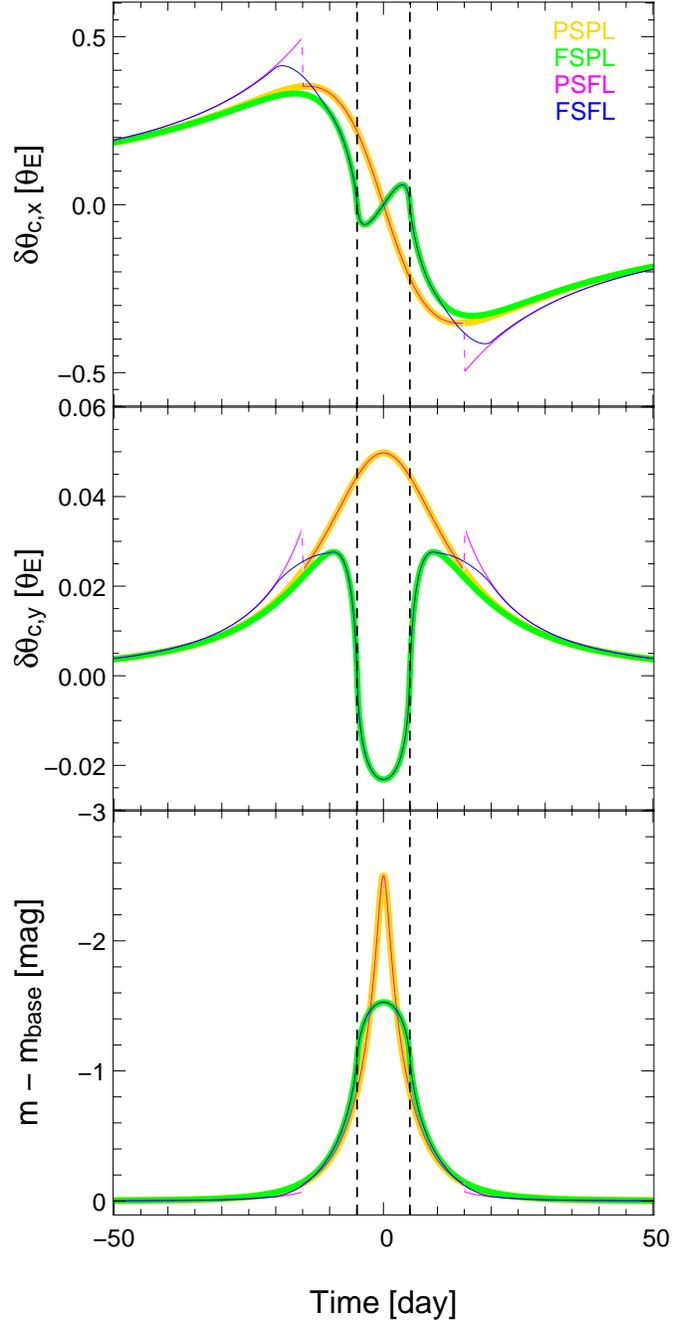}
  \caption{Centroidal shifts decomposition and light curves of a microlensing event 
           assuming $u_0$ = 0.1 $\AERR$, source radius = 0.5 $\AERR$ and lens radius 
           = 0.5 $\AERR$. We show the trajectory in the x- and y-direction (as in Fig.~\ref{fig.u0}) 
           of a PSPL (in yellow), FSPL (in green), PSFL (in red), and FSFL 
           (in blue). We also show the light curve with the magnitude variation relative to the 
           baseline ($m_{base}$). The vertical dashed line indicates the time when $u_0$ = $\RS$. 
           Note that the discontinuous part of the trajectory in a PSFL event is indicated by 
           the dashed red line. }
   \label{fig.compare_FSFL}
\end{figure}
\begin{equation}
  \begin{array}{l}
    \delta \mbox{\boldmath $\theta$}_{c,\mathrm{FSFL}} = \\
    \frac{\int_0^{2\pi} \int_0^{\RS} \left[A_+ \mbox{\boldmath $\theta$}_+ \Theta(\theta_+ - \RL)  + A_- \mbox{\boldmath $\theta$}_- \Theta(\theta_- - \RL)  \right] \LLD rdrd\phi }{\int_0^{2\pi} \int_0^{\RS}\left[A_+ \Theta(\theta_+ - \RL)  + A_- \Theta(\theta_- - \RL)  \right] \LLD rdrd\phi} - \mbox{\boldmath $u$}.
  \end{array}
  \label{eq.FSFL}
\end{equation}

The result is again shown in Fig.~\ref{fig.AM} 
along with the light curve. The trajectory first follows the PSFL trajectory at 
larger $u$, but, instead of a discontinuous jump, the FS effects is now bending the 
trajectory towards the FSPL trajectory until it fully becomes the FSPL 
trajectory at small $u$. 

To illustrate the importance of simultaneously including the finiteness of both the lens and the source, we 
compare $\delta \theta_{c,x} $ and $\delta \theta_{c,y} $ for the cases of PSPL, FSPL, PSFL, and FSFL in 
Fig.~\ref{fig.compare_FSFL}. When the size of the lens and the source are both negligible, it is clear that 
one would overestimate $\delta$\mbox{\boldmath $\theta$}$_c$ by adopting the PS 
approximation. On the other hand, taking the PL assumption would underestimate the value of 
$\delta$\mbox{\boldmath $\theta$}$_c$. Another important point is that in the FSFL scenario, one cannot determine 
the lens size by measuring the discontinuities in the trajectory presented by \cite{2003ApJ...595..418T} because the FS
effect makes the trajecotrty continuous. One thus needs to use equation~(\ref{eq.FSFL}) for deriving both $\RL$ and $\RS$.

\begin{figure}
  \includegraphics[scale=0.66]{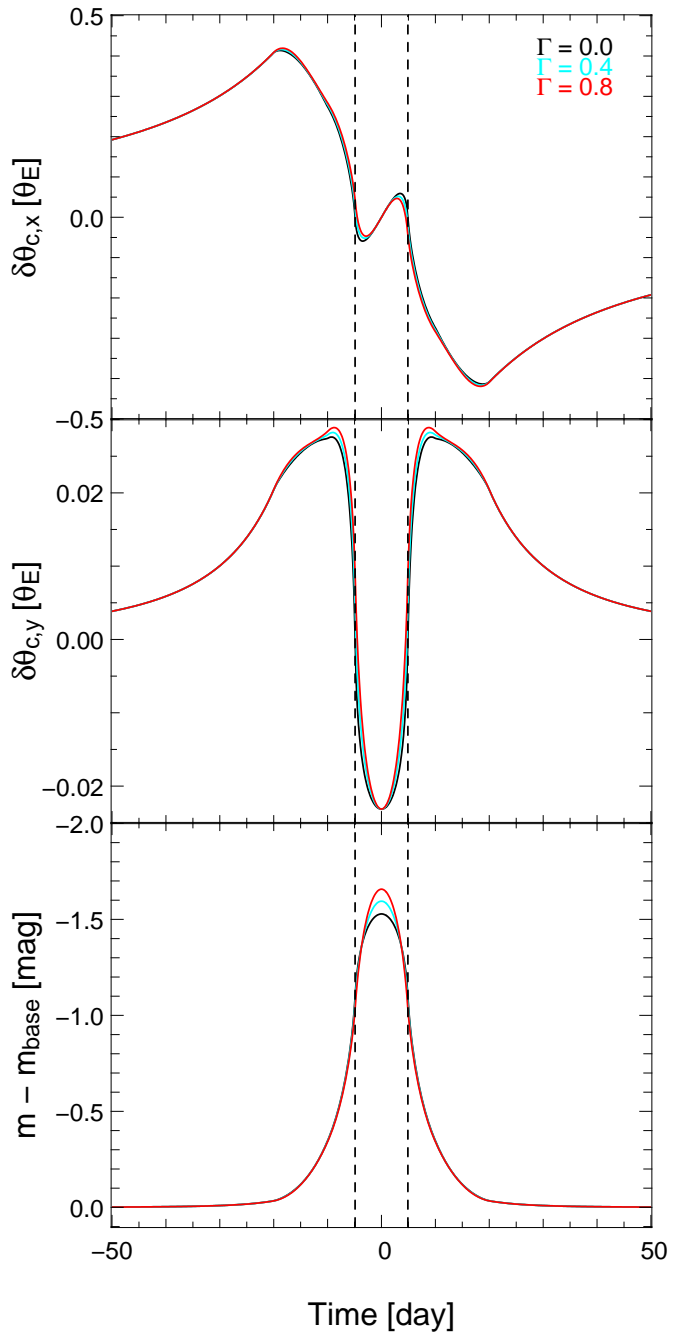}
  \caption{Centroidal shifts decomposition and light curves of a microlensing event 
           assuming $u_0$ = 0.1 $\AERR$, source radius = 0.5 $\AERR$ and lens radius 
           = 0.5 $\AERR$. We show the trajectory in the x- and y-direction (as in Fig.~\ref{fig.u0}) 
           of a uniform brightness source (in black), a limb-darkened source with $\LDC$ = 0.4 
           (in cyan) and $\LDC$ = 0.8 (in red). We also show the light curve with the 
           magnitude variation relative to the baseline ($m_{base}$). The vertical dashed line 
           indicates the time when $u_0$ = $\RS$.}
   \label{fig.compare_LD}
\end{figure}

We also show how the limb darkening changes the centroidal shift on top of a FSFL event 
(see Fig.~\ref{fig.compare_LD}) assuming different values of the limb-darkening coefficient $\LDC$.
In general, the limb-darkening only slightly modifies the astrometric trajectory. The FS 
effects and the limb darkening are most prominent when the lens transits the surface of the source, 
as indicated in Figs~\ref{fig.compare_FSFL} and \ref{fig.compare_LD}.

By fitting the centroidal shifts and/or the light curve as presented in Figs~\ref{fig.compare_FSFL} and 
\ref{fig.compare_LD}, one is able to constrain the value of $\RS$, $\RL$ and the limb-darkening 
coefficient $\LDC$. Events exhibiting FS effects have been detected photometrically 
\citep[e.g.][]{1997ApJ...491..436A, 2004ApJ...617.1307J, 2004ApJ...603..139Y, 2006A&A...460..277C, 
2009A&A...508..467B, 2009ApJ...703.2082Y, 2009arXiv0912.2312Z, 2010A&A...518A..51F}, and the
information of $\RS$ and $\LDC$ has been retrieved by fitting the light curve. Although it is hard to tell 
the difference between the FS and PS light curve by eye inspection, including the 
FS effects actually dramatically reduces the $\chi^2$ value for the best-fitted parameters. 
One can further fit the limb-darkening coefficient in different wavelength on top of the FS effects 
if multiwavelength observations are available. Practically, the limb darkening is relevant when
one wants to simultaneously fit photometric observations from different bands. 
However, one cannot measure the value of $\AERR$ directly from the light curve and thus the information 
of the actual source size is unknown. \cite{2000ApJ...534..894A} suggested to deal with this problem in the other way around. 
That is, given the colour information of the source by photometric observation, one can apply the 
relation between the colour and the surface brightness to obtain the actual source size if the stellar type 
of the source is known from the spectroscopic observation. Then the value of $\AERR$ can be calculated by 
$\AERR$:=$\AS / \RS$. Constraints on the lens mass and distance are also 
possible given the information of microlens parallax $\pi_{\mathrm{E}}$. However, inferring $\AERR$ 
photometrically from the source size is achievable only if the FS effect can be seen in 
the light curve, which is the case only when the lens transits the surface of the source, as discussed 
by \cite{1994ApJ...421L..71G}.

The advantage of astrometric microlensing is that the size of the astrometric signal is proportional to the 
value of $\AERR$. This means one can potentially measure $\AERR$ for every single 
event even if the FS effects in the light curve are not prominent. Given the information of $\AERR$,  
the actual size of the lens and the source are $\RS$ and $\RL$ multiplied by $\AERR$. 
It is also possible to compare the source size derived from the astrometric microlensing and from 
the colour to surface brightness relation.

\begin{figure*}
  \centering
  \includegraphics*[scale=0.8]{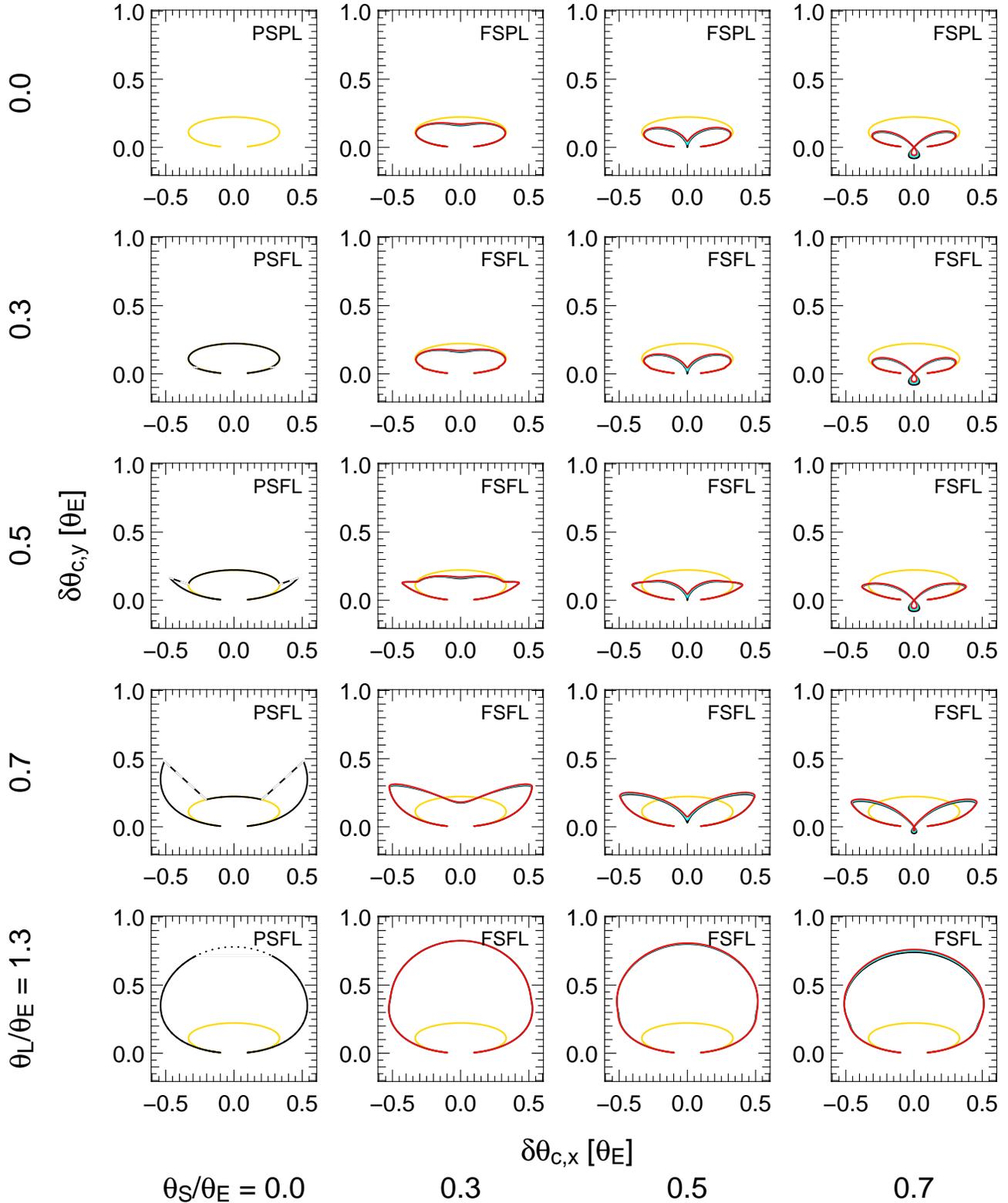}
  \caption{Centroid shifts in the presence of FS and FL effects for a dark lens. 
           We show the examples for a PSPL (in yellow), FSFL 
           with a uniform surface brightness source (in black) and with a 
           limb-darkened source (in cyan: $\LDC$ = 0.4 and in red: $\LDC$ = 0.8) 
           assuming $u_0$ = 0.5 $\AERR$ and with the sizes of the source and the lens 
           varying from 0.3 $\AERR$ to 1.3 $\AERR$. The dashed lines in the 
           plots with $\AS/\AERR$ = 0.0 show the discontinuity in the 
           trajectories for the PSFL cases. The dotted line in the PSFL with 
           $\AL / \AERR$ = 1.3 
           indicates the trajectory when the lens totally obscures both the plus- and minus-image.}
  \label{fig.FSFL}
\end{figure*}

We show the FSFL effects for a source of 
a uniform and a limb-darkend surface brightness profile with different source and lens sizes in 
Fig.~\ref{fig.FSFL} to illustrate how the combination of FS and FL 
influence the centroidal trajectory. The upper row of Fig.~\ref{fig.FSFL} gives the cases of 
PL approximation, which are comparable to the results of \cite{1998MNRAS.300.1041M}. The left-hand column of 
Fig.~\ref{fig.FSFL} shows the cases of PS approximation comparable to the results by \cite{2003ApJ...595..418T}. 

\section{The Luminous-Lens effects}
There are microlensing systems where both the source and the lens are resolved by \textit{HST} 
\citep[][]{2001Natur.414..617A, 2007ApJ...671..420K}. 
This indicates that the lens might be a luminous foreground star and thus perturbs the light centroid 
during the course of microlensing.  

Let us now consider the case where the light contribution from the lens is not negligible and 
start with the simple PSPL case. When the source is lensed by a 
point luminous lens (PSPLL), the centroid becomes the sum of the position multiplied by the flux of the 
two images and the lens over the total one. 
Furthermore, the reference point for the centroid shift is no longer the source centre, but the 
flux centre between the unlensed source and the lens instead 
\citep{1998ApJ...502..538B, 1999ApJ...511..569J, 2000ApJ...534..213D}

\begin{equation}
\delta \mbox{\boldmath $\theta$}_{c,\mathrm{PSPLL}} = \frac{A_+ \mbox{\boldmath $\theta$}_+ + A_- \mbox{\boldmath $\theta$}_- + \fls \mbox{\boldmath $\AL$}}{ A_+ + A_- + \fls } - \frac{\mbox{\boldmath $u$}}{1+ \fls},
\label{eq.PSPlL}
\end{equation}

where $\fls = f_{_{\mathrm{L}}} / f_{_{\mathrm{S}}}$ is the flux ratio between the lens and source and 
\mbox{\boldmath $\AL$} is the position of the lens on the lens plane. Here 
the $\fls$ \mbox{\boldmath $\AL$} term vanishes benefiting from the advantage of 
the lens-centred coordinates. For the case of FSFL, one just needs 
to modify the first part of equation~(\ref{eq.PSPlL}) by putting in the FL criteria 
of equation~(\ref{eq.PSFL}) and performing the integration over the source surface as 
equation~(\ref{eq.FSPL1D}) for a uniform brightness source or equation~(\ref{eq.FSPL}) for a more general 
source brightness profile, that is 
\begin{equation}
  \begin{array}{l}
    \delta \mbox{\boldmath $\theta$}_{c,\mathrm{FSFLL}} =\\
\\
    \frac{\int_0^{2\pi} \int_0^{\RS} \left[A_+ \mbox{\boldmath $\theta$}_+ \Theta(\theta_+-\RL) + A_- \mbox{\boldmath $\theta$}_- \Theta(\theta_--\RL) \right]\LLD rdrd\phi ~ + \fls \mbox{\boldmath $\theta$}_L}{\int_0^{2\pi} \int_o^{\RS} \left[A_+ \Theta(\theta_+-\RL) + A_- \Theta(\theta_--\RL) \right]\LLD rdrd\phi ~ + \fls} \\ 
\\
    - \frac{\mbox{\boldmath $u$}}{1+\fls}\quad.
    \label{eq.FSLFL}
  \end{array}
\end{equation}

\begin{figure}
  \includegraphics[scale=0.66]{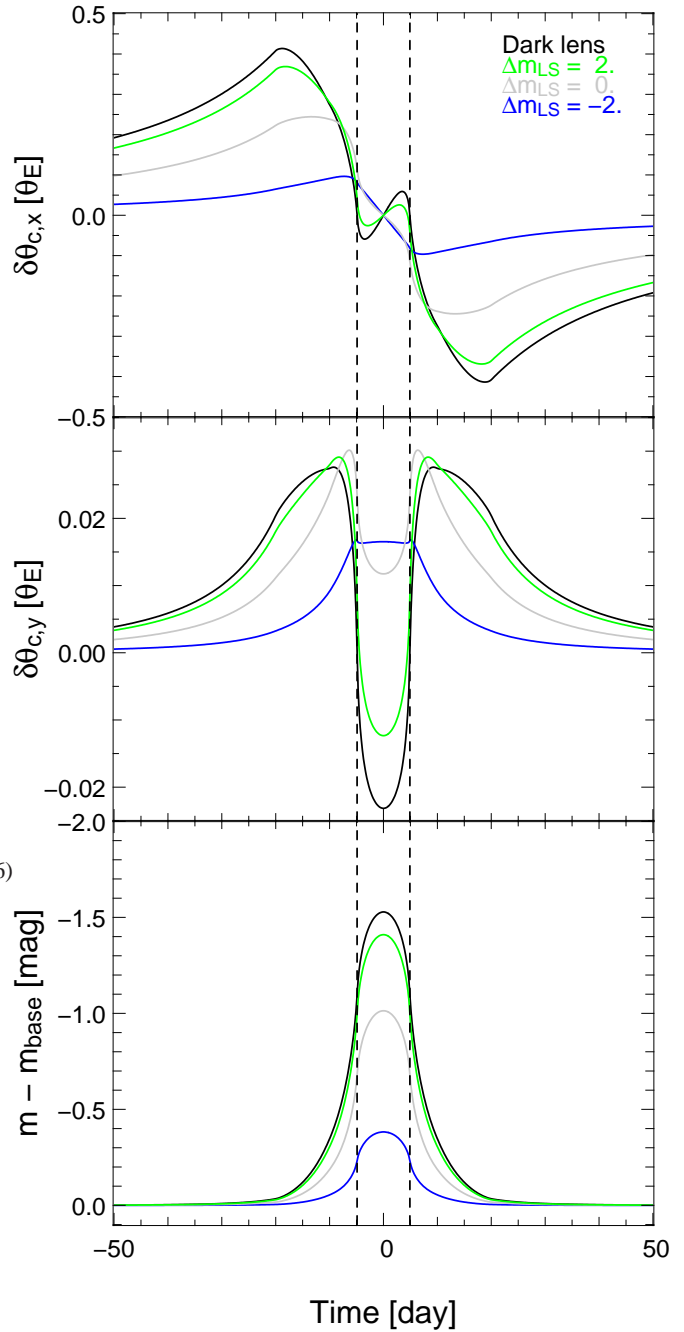}
  \caption{Centroidal shifts decomposition and light curves of a microlensing event 
           assuming $u_0$ = 0.1 $\AERR$, source radius = 0.5 $\AERR$ and lens radius 
           = 0.5 $\AERR$. We show the trajectory in the x- and y-direction (as in Fig.~\ref{fig.u0}) 
           of a bright lens with $\Delta m_{_{\mathrm{LS}}}$ = -2 (in blue), 0 (in gray), 
           2 (in green) and with a dark lens (in black). We also show the light curve with the 
           magnitude variation relative to the baseline ($m_{base}$). The vertical dashed line 
           indicates the time when $u_0$ = $\RS$.}
   \label{fig.compare_FSBL}
\end{figure}

Here we illustrate the influence of the luminous-lens effects on the centroidal shifts (and light curve) 
on top of a FSFL event in Fig.~\ref{fig.compare_FSBL}.
Since the limb-darkening only slightly modifies the trajectory
as shown in Fig.~\ref{fig.compare_LD}, we 
demonstrate the luminous-lens effects in the FSFL regime assuming a uniform brightness 
source in Fig.~\ref{fig.compare_FSBL}. We show the luminous-lens effects with various values for an 
apparent magnitude difference between the lens and source 

\begin{equation}
 \Delta m_{_{\mathrm{LS}}} = m_{_{\mathrm{L}}}-m_{_{\mathrm{S}}} = -2.5\mathrm{log}_{10}(\fls).
\end{equation}

\begin{figure*}
  \centering
  \includegraphics*[scale=0.8]{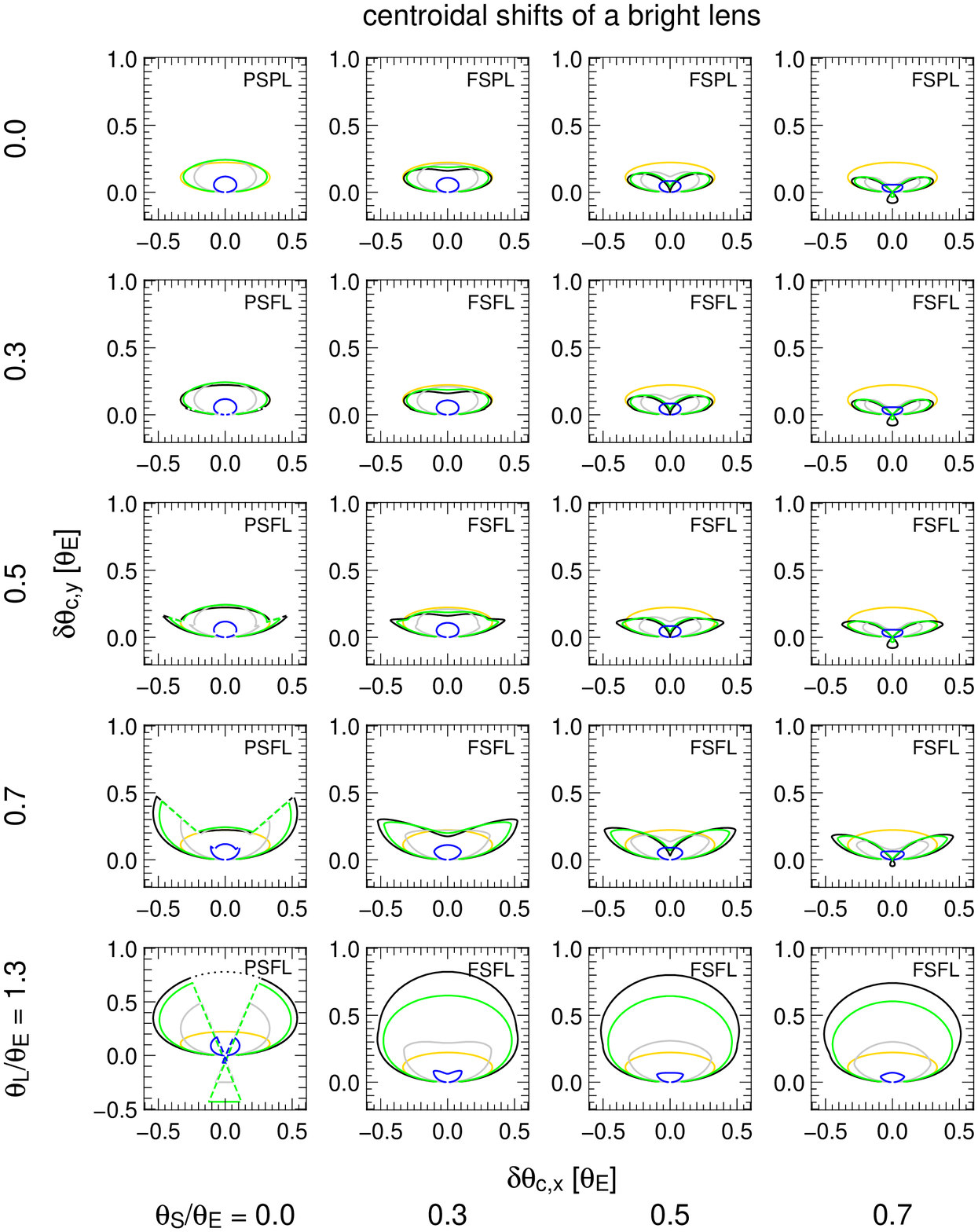}
  \caption{Centroid shifts in the presence of FS and FL effects for a luminous 
           lens. We show the examples for a  PSPL (in yellow), 
           FSFL with $\Delta m_{_{\mathrm{LS}}}$ = -2 (in blue), 0 (in gray), 
           2 (in green) and with a dark lens (in black) assuming $u_0$ = 0.5 
           $\AERR$ and with the sizes of the source and the lens varying from 0.3 
           $\AERR$ to 1.3 $\AERR$. The dashed lines in the plots with 
           $\AS/\AERR$ = 0.0 show the discontinuity in the trajectories 
           for the PSFL cases. The dotted lines in the PSFL with $\AL / \AERR$ = 1.3 
           indicates the trajectory when the lens totally obscures both the plus- and minus-image.}
  \label{fig.FSBL}
\end{figure*}

For illustrational purpose, we show the luminous-lens effects on top of the 
FSFL for different sizes of the lens and the source in Fig.~\ref{fig.FSBL}.
When the lens is getting brighter, the trajectory becomes smaller and rounder. The signal of centroidal 
shift is thus reduced for a source blended by a luminous lens.
The case of a PSPL events with luminous lens in Fig.~\ref{fig.FSBL} (upper left-hand corner) 
is comparable to the results of \cite{1999ApJ...511..569J}. Note that for the PSFL when $\RL$ = 1.3 (lower left-hand corner), 
the trajectory vanishes when $\RL > \theta_+$ for the dark lens case (black dotted line), but follows 
the trajectory of the lens for luminous-lens cases.

Since equation~(\ref{eq.FSLFL}) gives us the full consideration of the FSFL effects with the brightness of the source and 
lens (note that we only need to consider the flux ratio between the lens and the source, so the limb-darkening 
effects of the lens does not need to be taken into account), one is able to derive the information of 
$\RS$, $\LDC$, $\RL$, and $\fls$ by fitting the centroidal shifts. In principle, one can fit both the centroidal 
shifts and the light curve, to utilize both the astrometric and photometric information and thus to have a better 
constrain on the events parameters in equation~(\ref{eq.FSLFL}). Once the aforementioned parameters are all well 
determined, we can use the value of $\AERR$ to derive both the size of the lens and the source. 
We can also derive the mass of the lens as shown in equation~(\ref{eq.M}) given the information of the microlens 
parallax $\pi_{\mathrm{E}}$. The distance to the lens is also available if the distance to the source is well known, 
e.g. if the source is located in the Galactic bulge or Magellanic Clouds, which is often the case for 
the current microlensing surveys. In these cases, we are able to estimate the physical parameters of the whole 
microlensing systems.

\begin{figure*}
  \centering
  \includegraphics*[scale=0.5]{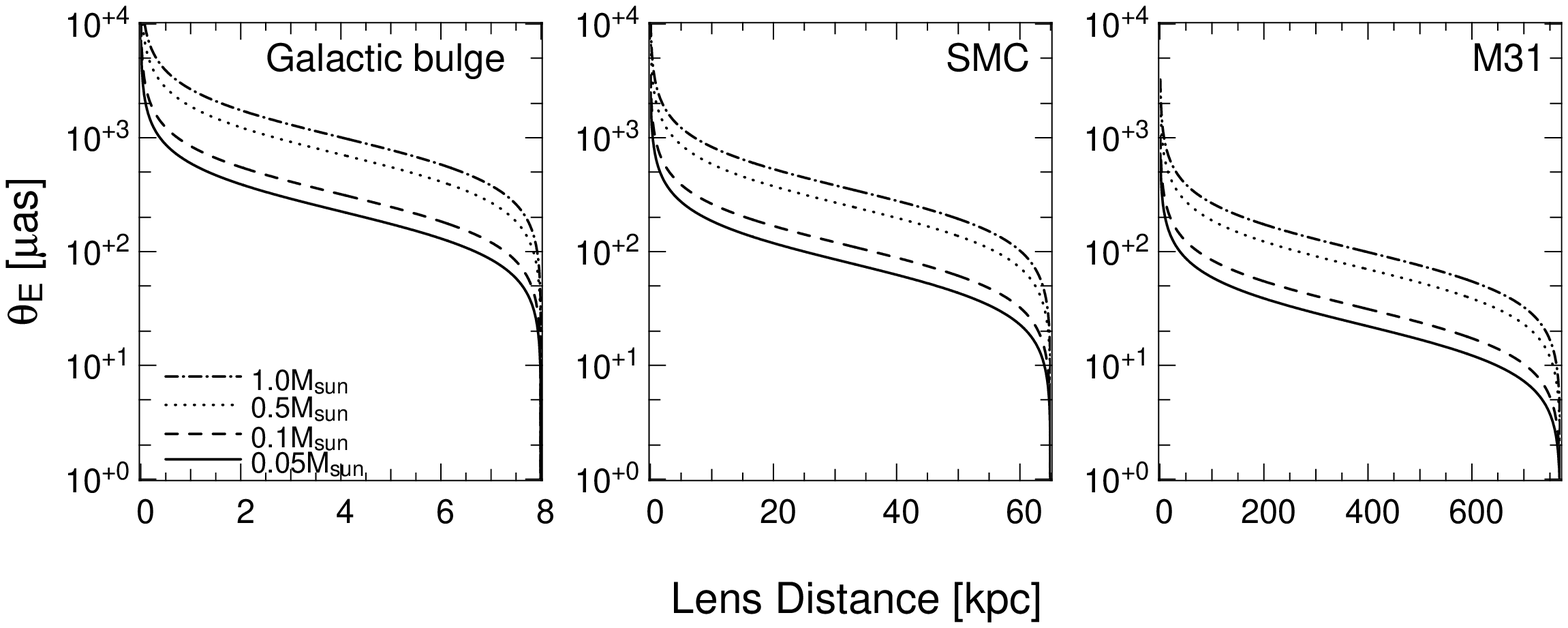}
  \includegraphics*[scale=0.5]{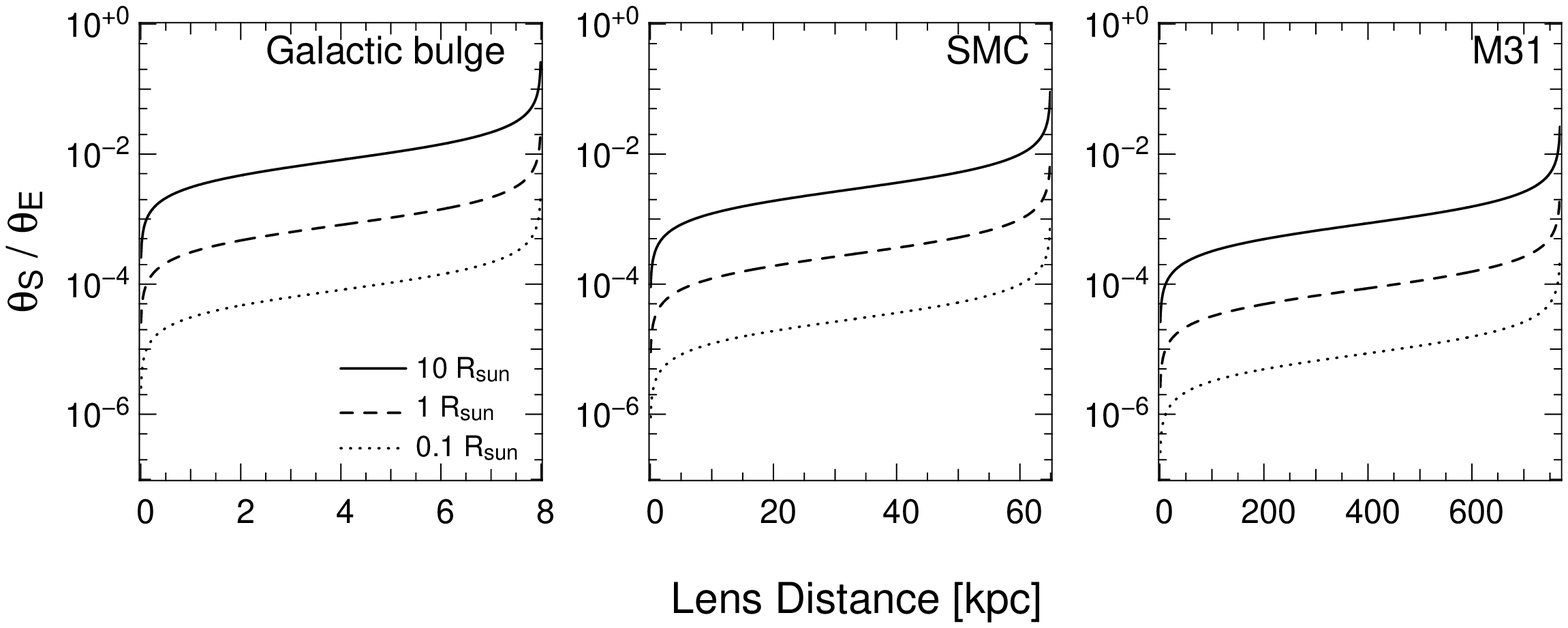}
  \includegraphics*[scale=0.5]{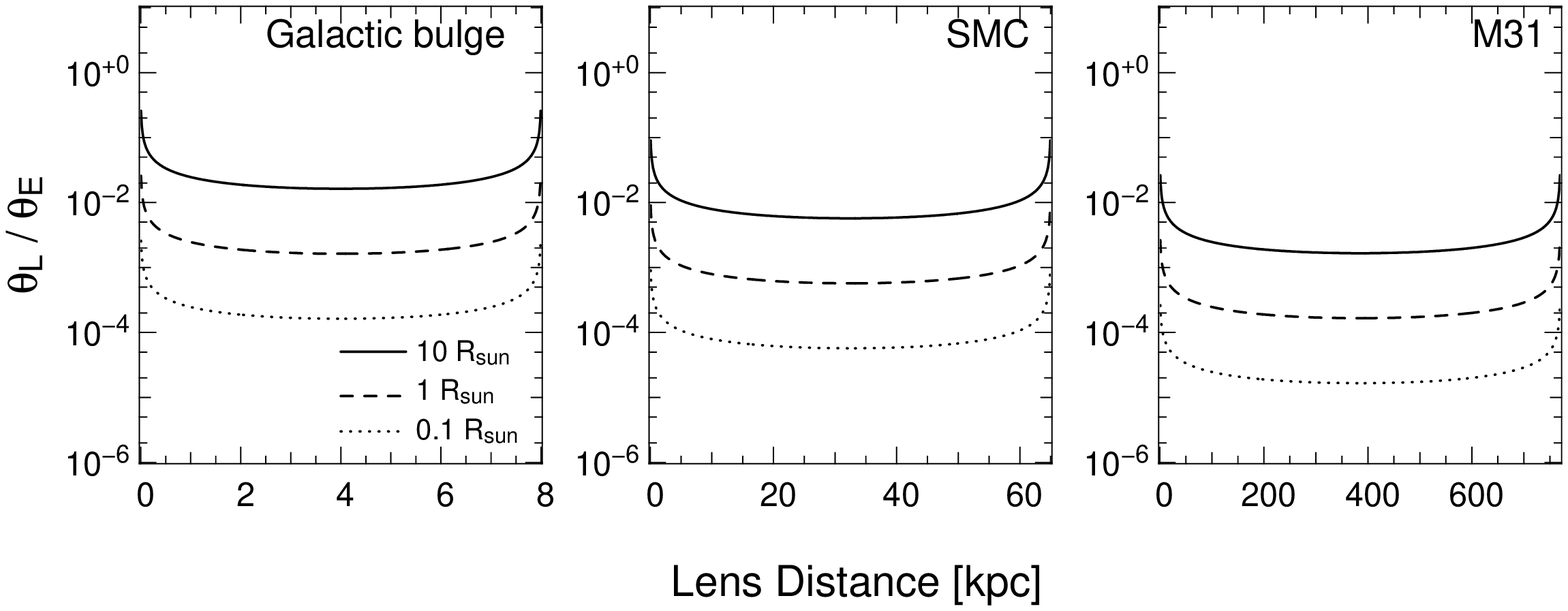}
  \caption{First row: $\AERR$ as a function of $\Dol$ assuming the source 
           located in the Galactic bulge ($\Dos=8$ kpc), SMC ($\Dos=65$ kpc), and 
           M31 ($\Dos=770$ kpc). Second and third row: $\AS / \AERR$ and 
           $\AL / \AERR$ as a function of the lens distance assuming $\ML 
           = 0.5 M_{\odot}$.}
  \label{fig.thetaE}
\end{figure*}

\section{Observational Feasibility}

In this section we consider the astrometric events towards the Galactic bulge, SMC and M31 assuming 
$\Dos$ = 8, 65, and 770 kpc, respectively. We substitute $\Dol / \Dos $ = $x$ into 
equation~(\ref{eq.tE}), which then becomes 

\begin{equation}
\AERR = \sqrt{\frac{4G\ML}{c^2 \Dos}\left(\frac{1}{x}-1 \right)}\quad .
\label{eq.thetaE_new}
\end{equation}

Therefore, $\AERR$ is smaller for a source located at larger distance and is smaller 
for larger lens distance given the same source location (see upper panels in Fig.~\ref{fig.thetaE}). 

Equation~(\ref{eq.thetaE_new}) also implies that the halo lensing events have larger 
Einstein radii than self-lensing events for a given lens mass. For instance, halo lensing events 
towards SMC with $\Dol$ = 15 kpc and $\ML$ = 1 $M_{\odot}$ will induce an astrometric signal 
with $\AERR$ = 645 $\mu$as, which is one order of magnitude larger than for 
self-lensing events (44 $\mu$as at $\Dol$ = 64 kpc). Thus we are able to distinguish halo and self-lensing events 
by the size of the astrometric ellipse.

The FS effects play an important role when $u_0 \le \RS$, which is often the case 
when $u_0 \ll$ 1 
\citep{1994ApJ...421L..71G}. 
However, such a configuration leads to a smaller centroidal shift 
(as shown in Fig.~\ref{fig.u0}) and is thus very challenging to distinguish between the PSPL and FSPL 
trajectories observationally.

FL effects are prominent when $\RL$ is close to and larger than unity 
(as shown in Fig.~\ref{fig.FSFL}).
We thus calculate $\RL$ by dividing the angular lens radius $\AL$ by 
$\AERR$ (see the lower panel in Fig.~\ref{fig.thetaE}). Because $\AL$ is 
proportional to $1/x$ while $\AERR$ is a function 
of $\sqrt{1/x - 1}$, $\RL$ is actually a function of $[x(1-x)]^{-1/2}$. We would expect to see the 
FL effects when the lens is located either close to the observer ($x \approx 0$) 
or to the source ($x \approx 1$). By equating $\AL$ to $\AERR$, we have
\begin{equation}
\frac{\Dol}{\Dos}\left(\Dos-\Dol \right) = \frac{R_{_{\mathrm{L}}}^2c^2}{4G\ML}.
\label{eq.calc_rL}
\end{equation}

\begin{figure*}
  \centering
  \includegraphics*[scale=0.45]{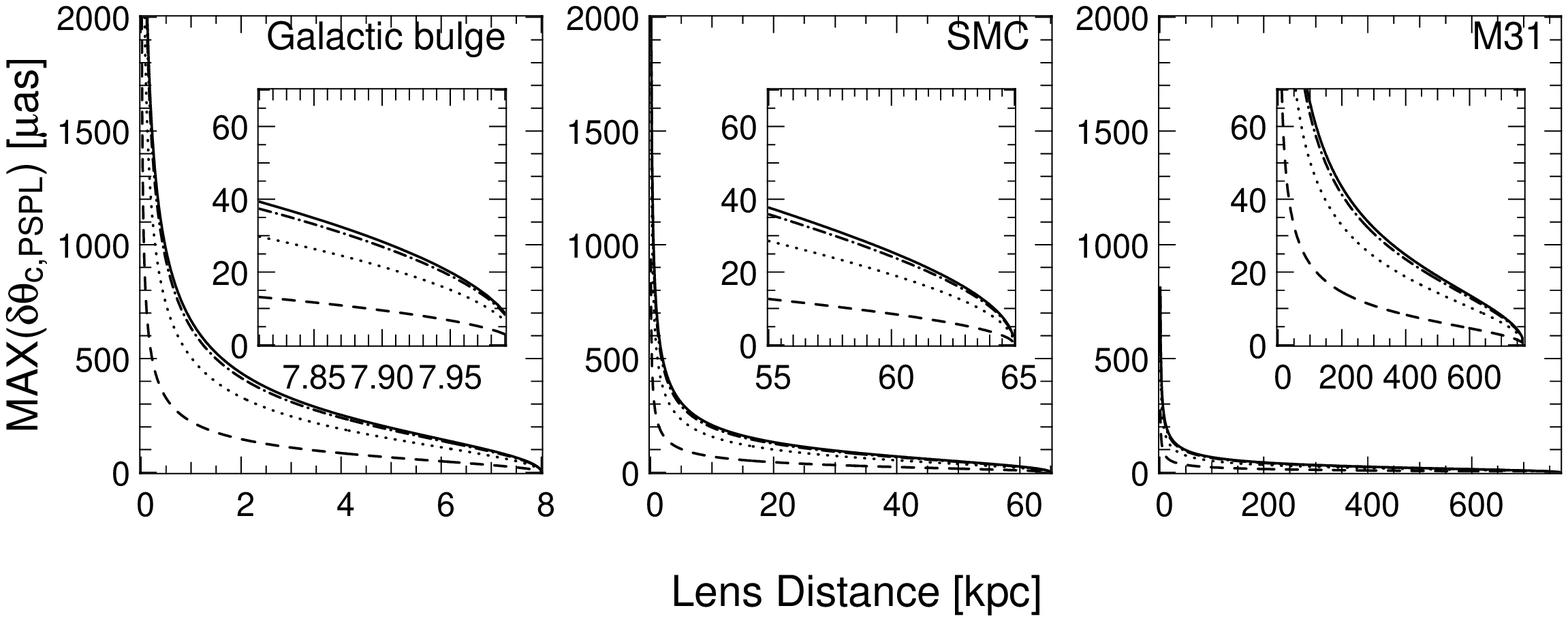}
  \includegraphics*[scale=0.45]{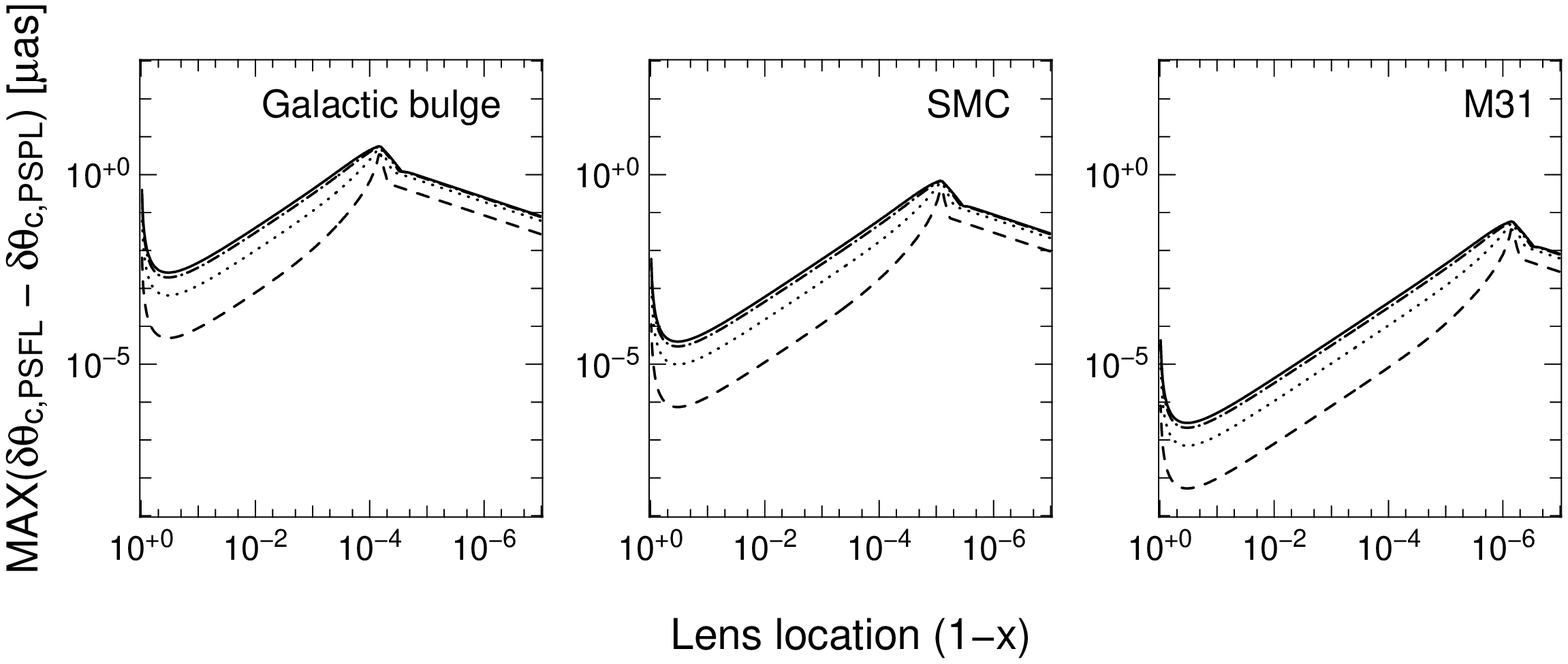}
  \includegraphics*[scale=0.45]{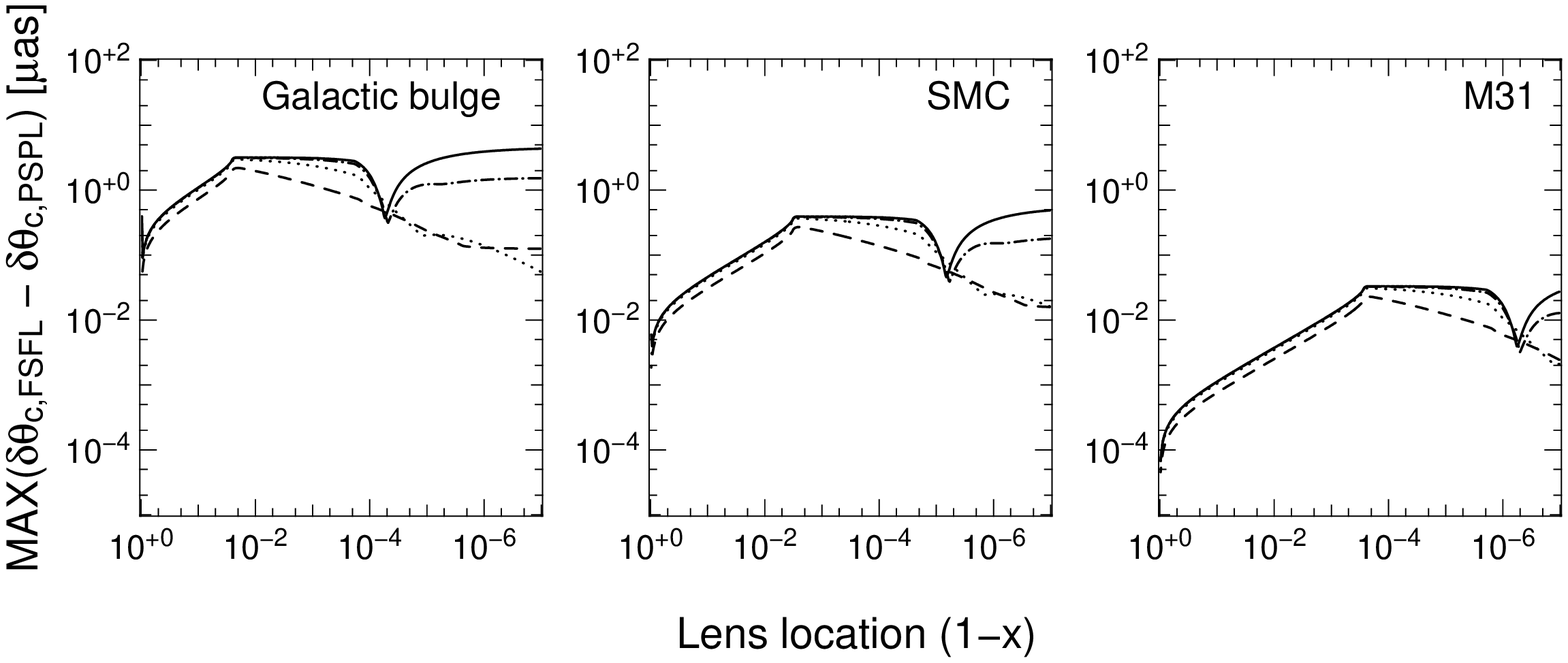}
  \includegraphics*[scale=0.45]{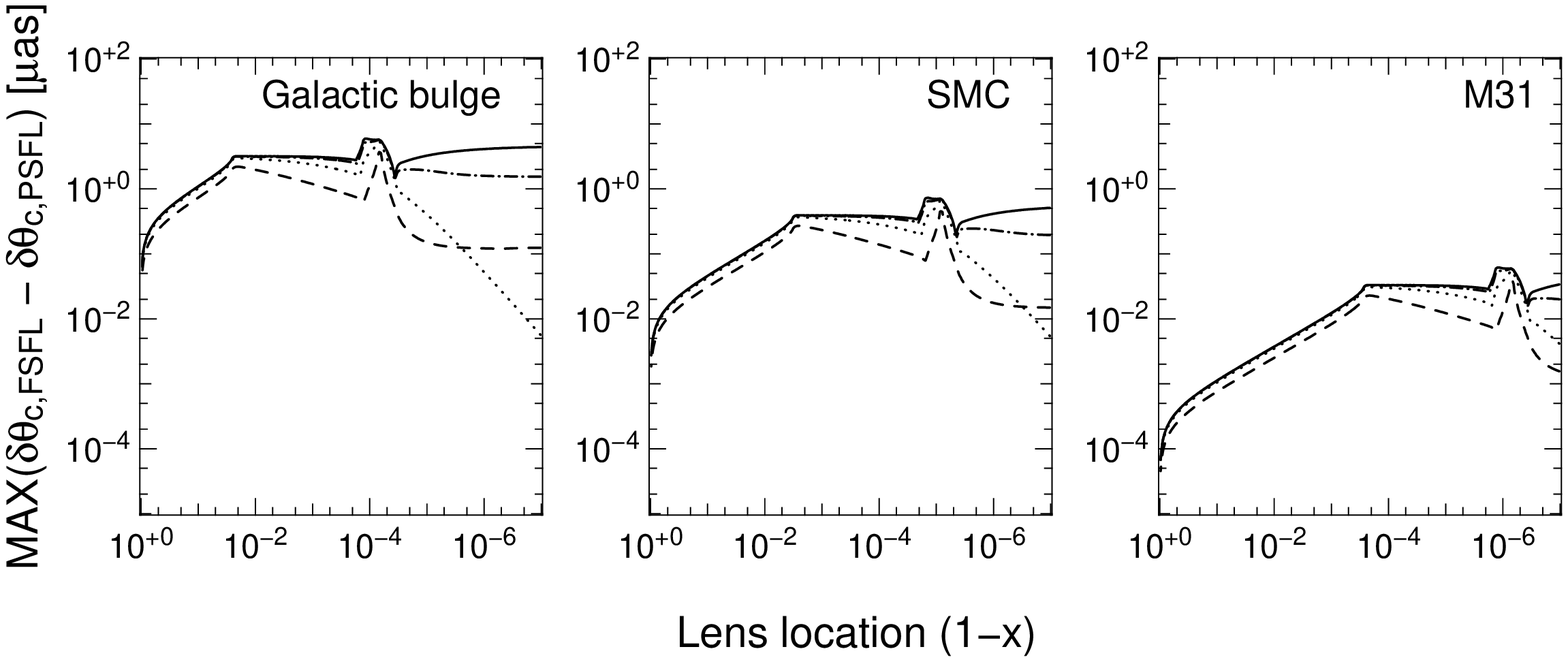}
  \caption{First row: maximum centroidal shifts as a function of $\Dol$ assuming the source located in the 
           Galactic bulge ($\Dos=8$ kpc), SMC ($\Dos=65$ kpc), and M31 ($\Dos=770$ kpc) 
           in PSPL. Second row: maximum deviation between the PSFL and PSPL trajectories as a function of the 
           lens location $(1-x)$, where $x = \Dol/\Dos$. 
           Third row: maximum deviation between the FSFL and PSPL trajectories.
           Fourth row: maximum deviation between the FSFL and PSFL trajectories.
           We assume $u_0$ = 0.05 $\AERR$, $t_{\mathrm{E}}$ = 10 days, 
           $\ML$ = 0.5 $M_{\odot}$, $R_{\mathrm{L}}$ = 10 $R_{\odot}$, and $R_{\mathrm{S}}$ = 10 $R_{\odot}$. 
           Here we show the cases of a luminous lens with $\Delta m_{_{\mathrm{LS}}}$ = 2 (in dashed), 
           0 (in dotted), -2 (in dash-dotted), and a dark lens (in solid).} 
  \label{fig.Max_thetaC}
\end{figure*}

\begin{figure*}
  \centering
  \includegraphics*[scale=0.4]{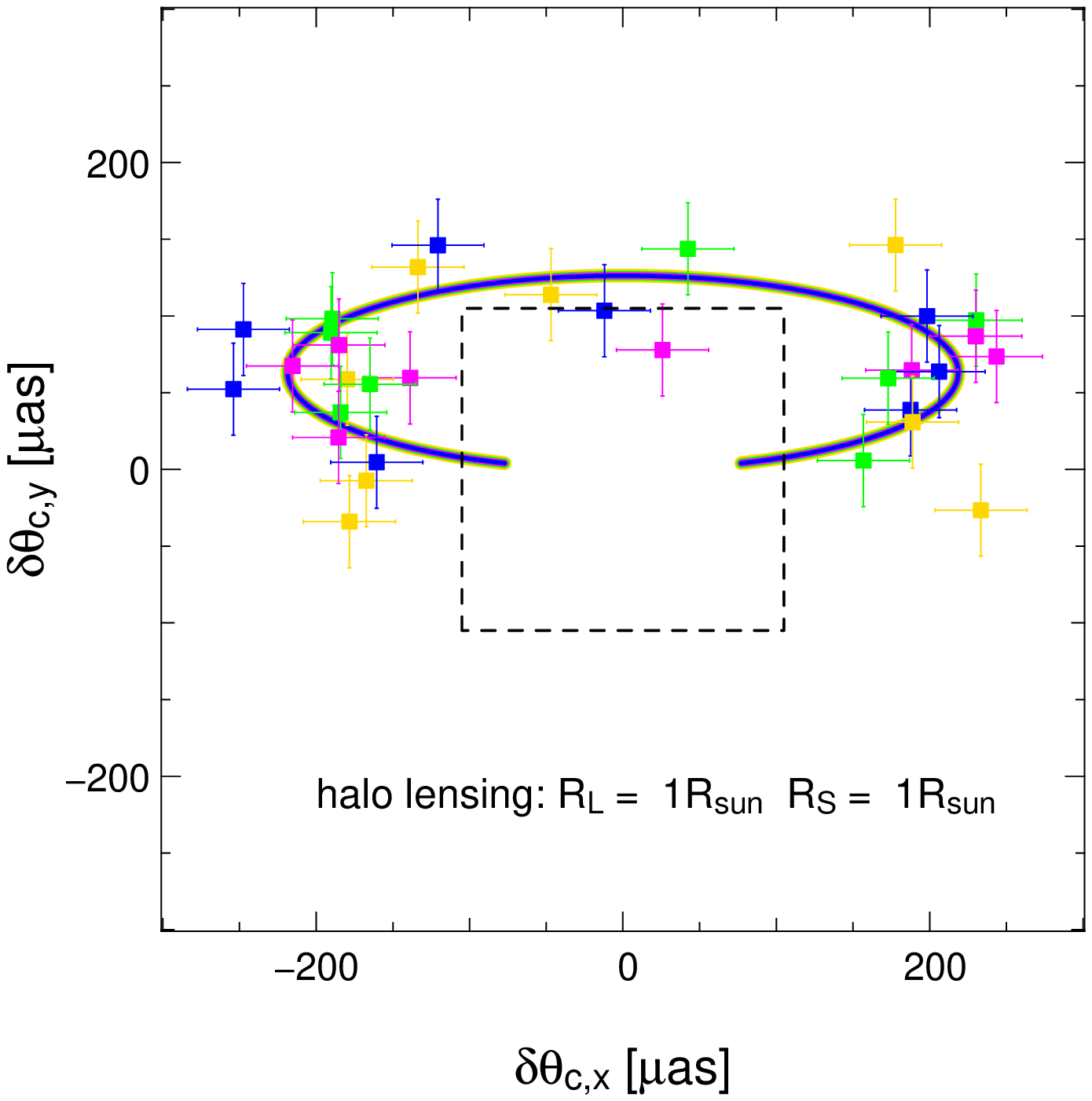}
  \includegraphics*[scale=0.4]{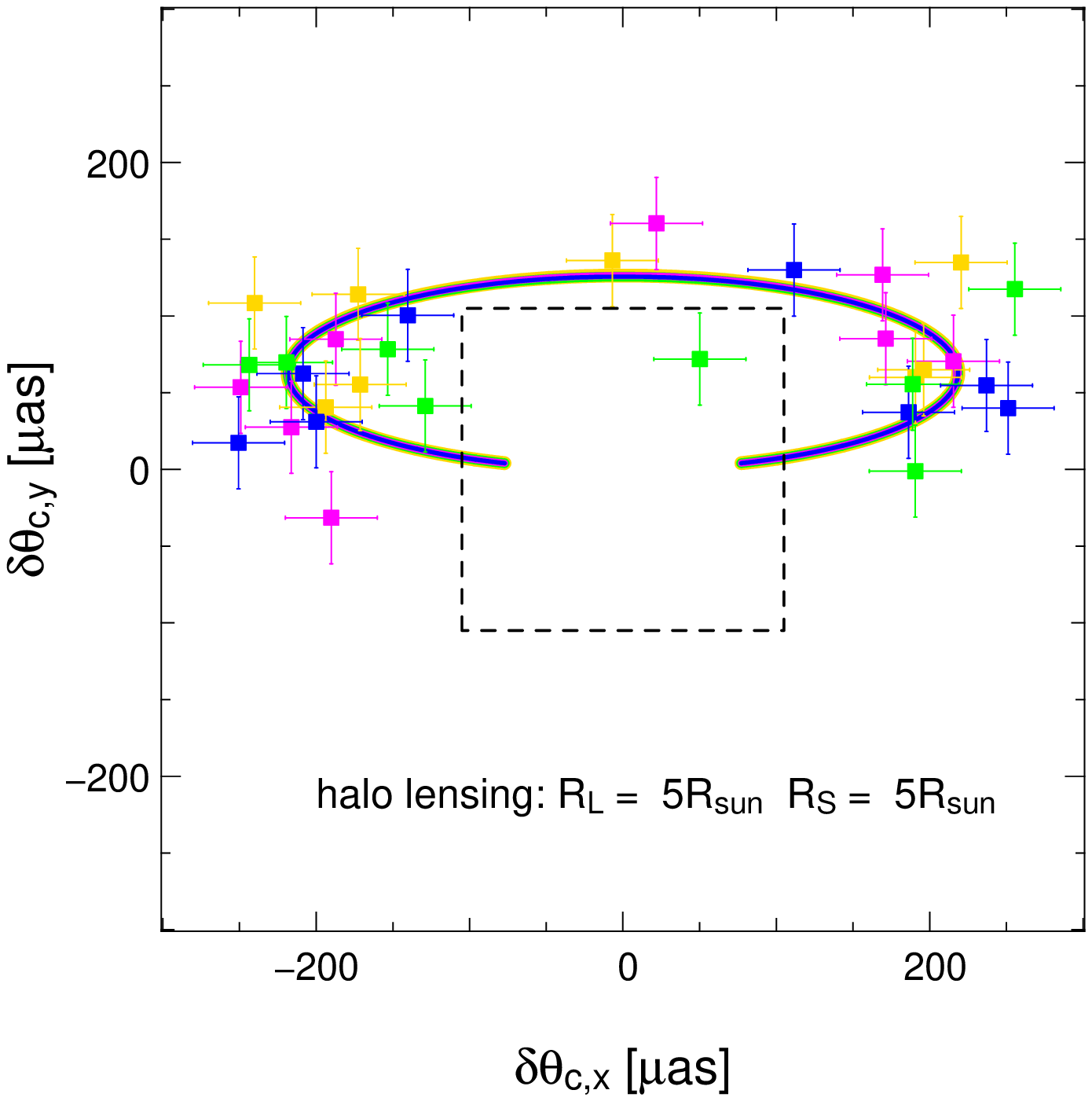}
  \includegraphics*[scale=0.4]{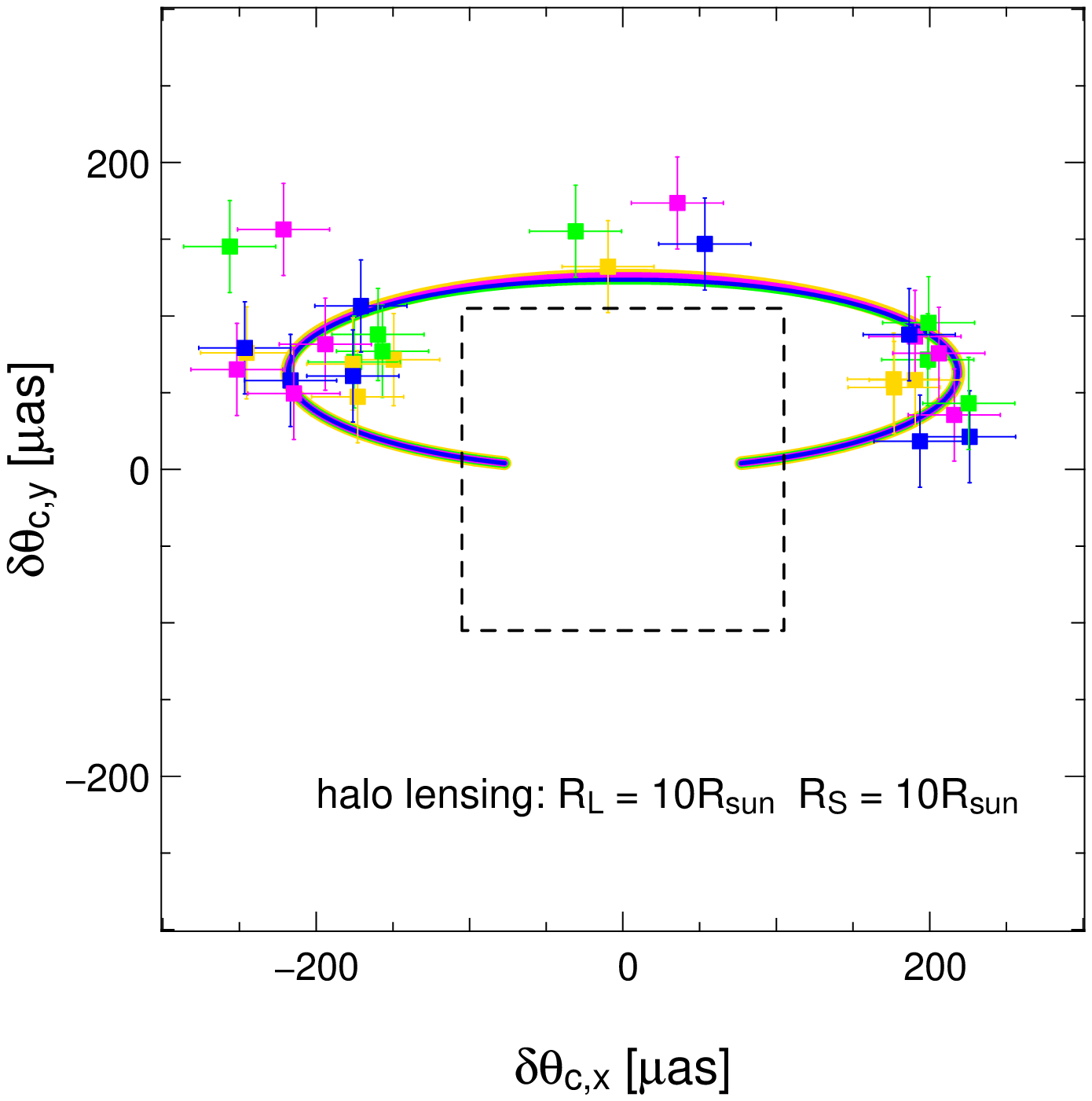}
  \includegraphics*[scale=0.4]{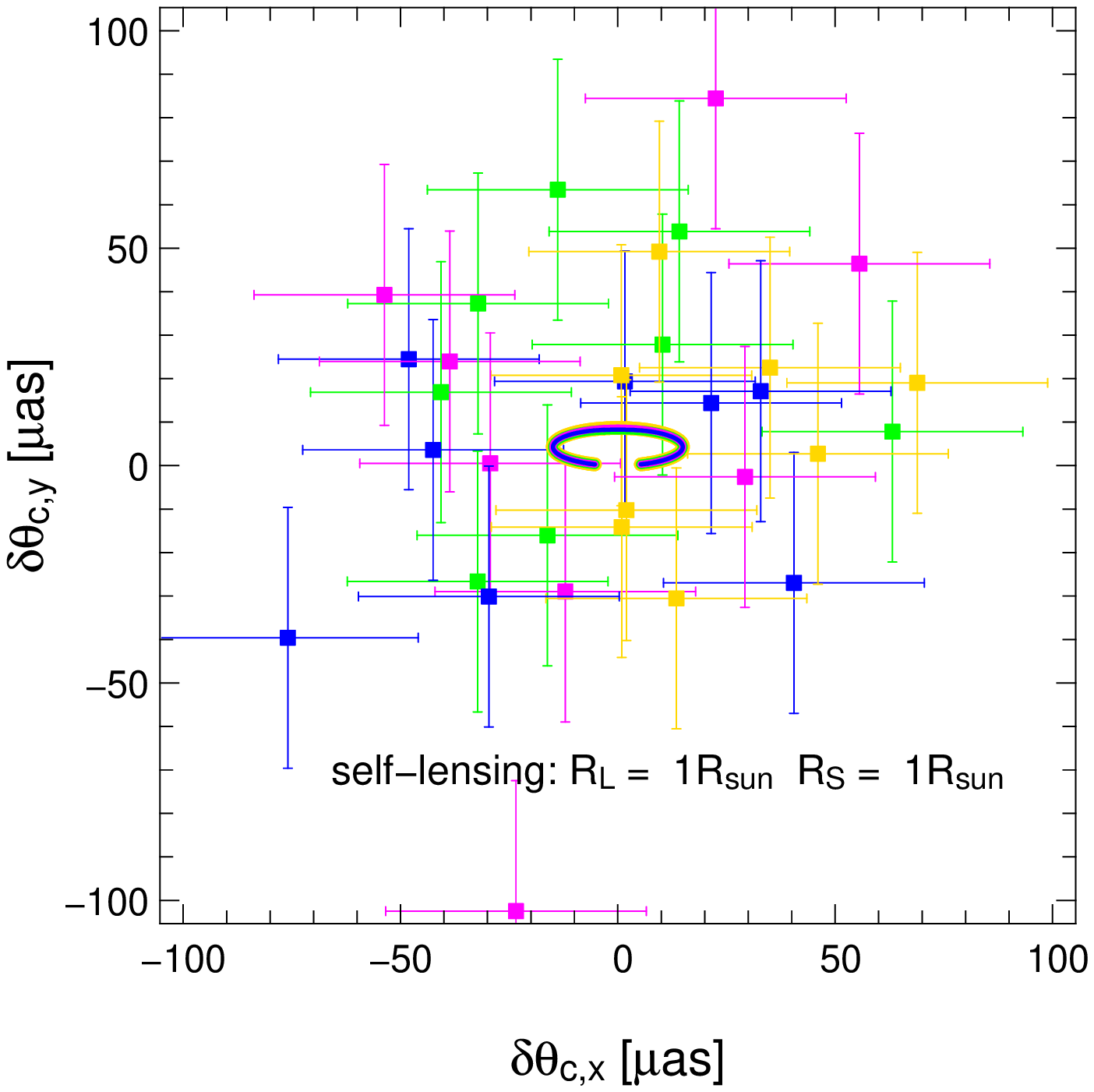}
  \includegraphics*[scale=0.4]{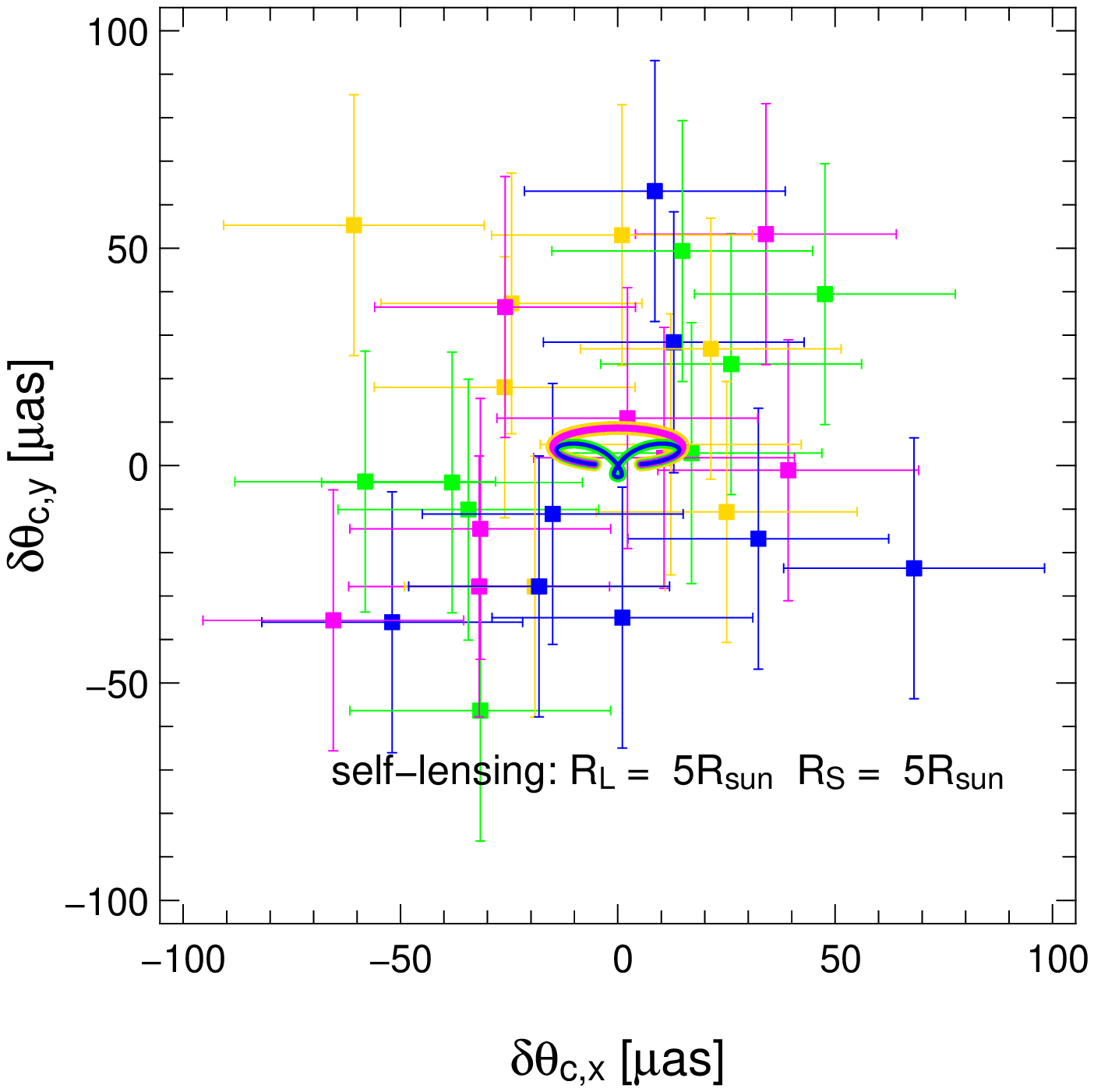}
  \includegraphics*[scale=0.4]{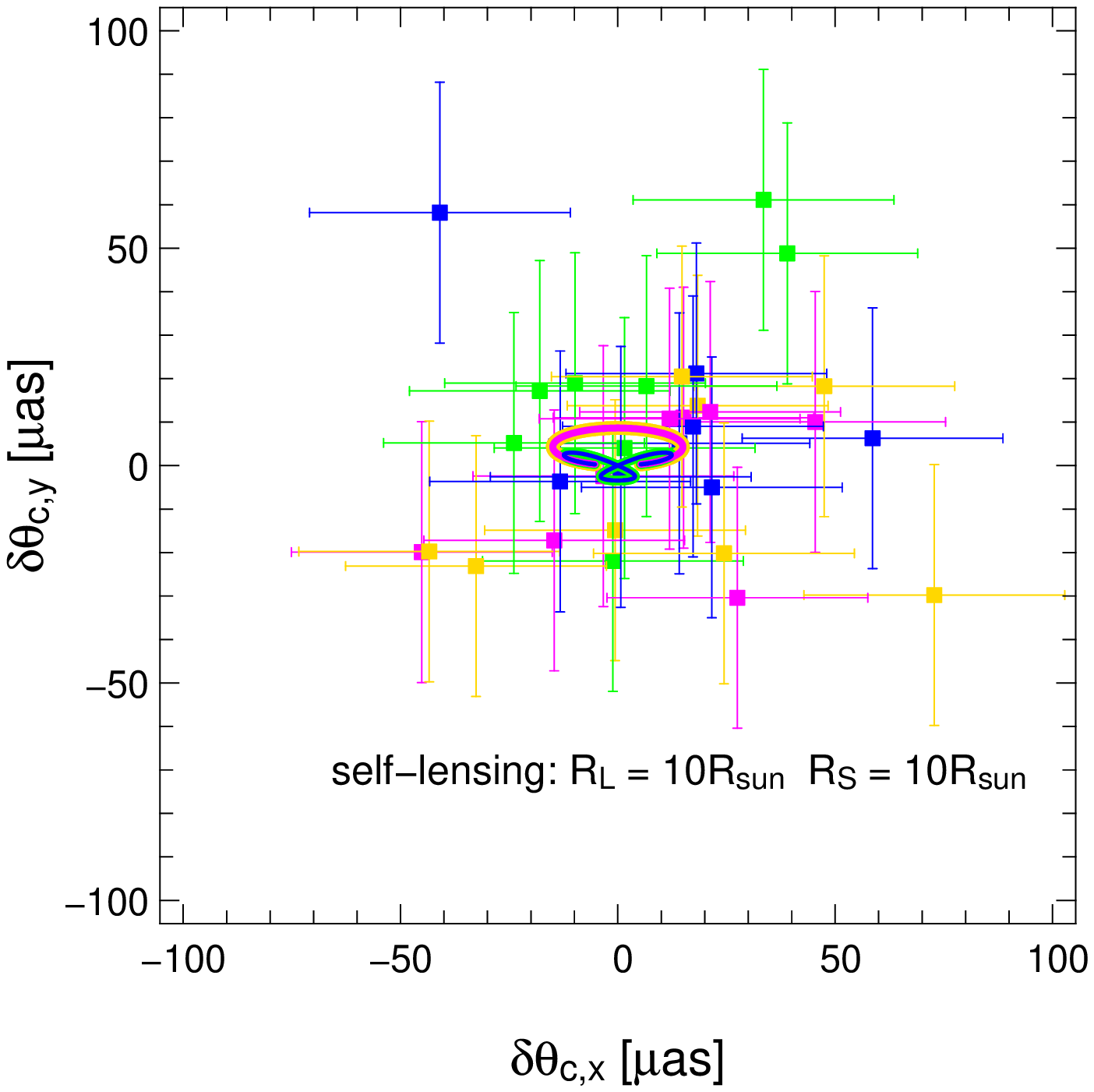}
  \caption{Astrometric trajectory of the SMC microlensing event MACHO-97-SMC-1 with 
           $t_{\mathrm{E}}$ = 123.5 days and $u_0$ = 0.426824 $\AERR$. We show the case 
           when the lens is located in the Galactic halo (upper panels) and in SMC (lower panels). 
           We assume the lens is 1 $M_{\odot}$ and the size of the lens and the source are assumed 
           to be 1, 5, and 10 $R_{\odot}$, respectively. The color convention is the same as 
           Fig.~\ref{fig.AM}. We calculate the theoretical trajectory within a time interval of $t_0 \pm$ 
           1000 days. We then simulate the measurements of \textit{SIM}, with a sampling rate of 
           every 90 days spanning for $t_0 \pm$ 1 year and 30-$\mu$as error in both x and y directions. 
           The dashed square in the halo lensing cases outlines the dimension showed in the 
           self-lensing regime.}
  \label{fig.SMC}
\end{figure*}

The left-hand side of this equation gives us the information on the location of the lens to have prominent 
FL effects \citep[]{2002ApJ...579..430A}. If the lens is very close to the observer such that 
$\Dol \ll \Dos$, equation~(\ref{eq.calc_rL}) gives us an upper limit of $\Dol$ so that for lenses beyond 
this value, the FL effect is not prominent. On the other hand, if the lens is very close 
to the source so that $\Dol \approx \Dos$, then equation~(\ref{eq.calc_rL}) actually gives us a maximum separation 
between the lens and the source in order to have non-negligible FL effects. The value of 
$\Dol(\Dos-\Dol)/\Dos$ for several astrophysical objects are given in Table~\ref{tab.calc_rL}.

\begin{table}
  \centering
  \caption{$\Dol(\Dos-\Dol)/\Dos$ for several astrophysical objects.}
  \begin{tabular}{@{}lrrr@{}}
    \hline
    Name         &    Radius ($R_{\odot}$) &  Mass ($M_{\odot})$    &        $\frac{\Dol}{\Dos}(\Dos-\Dol)$ \\
    \hline
    Sun          &                    1. &                  1. &      551 AU \\
    Jupiter      &                   0.1 &               0.001 &      5.5$\times 10^{3}$ AU \\
    Earth        &                 0.009 &  3$\times 10^{-6}$  &      1.5$\times 10^{4}$ AU \\
    Brown dwarf  &                   0.1 &                0.05 &        110 AU \\
    White dwarf  &                 0.009 &                  1. &       0.045 AU \\
    Neutron star &   2.8$\times 10^{-5}$ &                  1.4 &    3.1$\times 10^{-7}$ AU \\
    Black hole   &   4.2$\times 10^{-5}$ &                  10. &    9.9$\times 10^{-8}$ AU \\
    Test case    &                   10. &                  0.5 &    1.1$\times 10^{5}$ AU \\
    \hline
  \end{tabular}
  \label{tab.calc_rL}
\end{table}

We then calculate the maximum centroid deviation versus lens distance for the cases of PSPL, PSFL, FSPL, and 
FSFL assuming that a source of 10 $R_{\odot}$ is amplified by a lens of 0.5 $M_{\odot}$ and 10 $R_{\odot}$ with the 
minimum lens-source separation projected onto the sky to be 0.05 $\AERR$. Because there are only small differences 
between PSPL, PSFL, FSPL, and FSFL, we only show the case of PSPL in Fig.~\ref{fig.Max_thetaC}.
To see how much the FL trajectory deviates from that of PL and the influence from FS, we further 
calculate the maximum difference between the PSPL, PSFL, and FSFL trajectories 
at a given time. The result is shown in Fig.~\ref{fig.Max_thetaC} for a lens with $\Delta m_{_{\mathrm{LS}}}$ = 
2, 0, -2 and a dark lens. The FSFL only shows small difference to that of PSFL and PSPL. The difference is less than 
10 $\mu$as for the case of Galactic bulge, and even smaller than 1 $\mu$as for the more distant source in the SMC and M31.  
This is because the FL effect is important only when the lens is extremely close to the observer or 
the source and the major difference between FSFL and PSPL or PSFL comes from the finiteness of the source.  
Since $u_0$ is larger than $\RS$ for most of the time, the FS effect only slightly changes the 
astrometric trajectory. Even when the lens is extremely close to the source (see Fig.~\ref{fig.Max_thetaC}), 
the already reduced $\AERR$ makes the difference so small that it is hardly observable.

In order to test if the astrometric signal is observable towards SMC, we simulate the astrometric
trajectory of MACHO-97-SMC-1 \citep{1997ApJ...491L..11A}. This event has baseline magnitude 
\textit{V} = 17.7, so it will take \textit{SIM} $\sim$ 3 hours to reach 30-$\mu$as accuracy 
\citep{2008SPIE.7013E.151G}. We thus simulate observations by \textit{SIM} assuming 
the measurement errors to be Gaussian distribution with $\sigma$ = 30 $\mu$as. 
We put the lens at a distance of 15 kpc and 64 kpc corresponding 
to the halo and self-lensing scenario towards SMC. We then assign a putative finite size of 1, 5, 
and 10 $R_{\odot}$ to the lens and the source. The mass of the lens is set to be 1 $M_{\odot}$. 
From Fig.~\ref{fig.SMC} we can see that if the lens is in the Galactic halo, we are able to detect 
the astrometric signal because of the very large $\AERR$. However, the finite size of the 
source and the lens is not revealed in such a close lens. On the other hand, the FS and FL effects 
are prominent in the self-lensing regime due to the small $\AERR$. 
But the astrometric trajectory is too small to be constrained by current instruments, not to mention 
to disentangle between the PSPL and FSPL or PSFL. Nevertheless, it is still possible to use the 
(non-)detection of the astrometric ellipse to infer if the lens is in the Galactic halo or it is a 
self-lensing event towards SMC.

We also considered the possibility to detect the astrometric trajectory from ground-based 
instruments such as \textit{PRIMA} for VLTI. \textit{PRIMA} can determine the astrometry to 
10-$\mu$as level in 30 minutes provided a reference star within 10 arcsec and a 200-m baseline (ATs mode). 
The goal of \textit{PRIMA} is to perform astrometric measurement for a target as faint as 18 (15) mag 
with UTs (ATs) provided a 13 (10) mag reference star in \textit{K} band  \citep{2008NewAR..52..199D}. 
There is a bright star (\textit{K} = 10.28) in the vicinity of MACHO-97-SMC-1 
(separated at 30.4 arcsec), so theoretically it would be possible to obtain 30-$\mu$as accuracy in 
astrometric measurements within one hour with the UTs (130 m baseline). However, for the two stars separated by 
20 arcsec, there is already 90\% reduction in the interferometric fringe visibility. Thus it would be very challenging to 
conduct such measurement.
It would be very difficult to routinely measure the astrometry towards SMC/LMC with 
\textit{PRIMA} because most of the single lens events in the Magellanic Clouds (14 out of 15, except 
MACHO-97-SMC-1) have sources fainter than 19 mag in \textit{V} 
(1 in \citealp{1997ApJ...491L..11A}; 12 in \citealp{2000ApJ...542..281A}; 
1 in \citealp{2007A&A...469..387T}, which is the same as \citealp{1997ApJ...491L..11A}, and 2 in 
\citealp{2009MNRAS.397.1228W}).

To perform astrometric measurements for microlensing events towards M31 is beyond the limit of 
both \textit{PRIMA} and \textit{SIM} since the sources in M31 are too faint 
(see e.g. Riffeser et. al., in preparation, and reference therein).

\section{Conclusion}
We have studied the astrometric aspects of microlensing by simultaneously including the FS 
and FL effects. Our results show that the astrometric signal is underestimated or overestimated by 
assuming PL or PS, respectively. While the FS effect is prominent when 
the lens transits the surface of the source, the FL effect is revealed when the lens is very 
close to the source, which would be in the self-lensing regime. In the context of the self-lensing scenario, 
where a background star is lensed by a foreground star, the light contribution from the 
lens is in general not negligible. We thus consider the luminous-lens scenario, which attenuates the signal of 
the centroidal displacement. 
Astrometric trajectories with a source located in the Galactic bulge, SMC, and M31 are discussed, 
which show that $\AERR$ of halo-lensing events is at least one order of magnitude 
larger than that of self-lensing in SMC and M31. Our results also indicate that the finiteness 
of the lens is more likely to be revealed in the self-lensing scenario towards distant source 
located in Magellanic Clouds or M31, although it is very difficult to distinguish between PL 
and FL with current instruments.

\section*{Acknowledgments}
We are very grateful to the anonymous referee for the useful comments. 
This work was supported by the DFG cluster of excellence `Origin and Structure of the Universe' 
(www.universe-cluster.de).


\begin{thebibliography}{99}
\bibitem[\protect\citeauthoryear{Agol}{2002}]{2002ApJ...579..430A} Agol E., 
2002, ApJ, 579, 430 
\bibitem[\protect\citeauthoryear{Albrow et al.}{2000}]{2000ApJ...534..894A} 
Albrow M.~D., et al., 2000, ApJ, 534, 894 
\bibitem[\protect\citeauthoryear{Alcock et al.}{1997a}]{1997ApJ...491L..11A} 
Alcock C., et al., 1997a, ApJ, 491, L11 
\bibitem[\protect\citeauthoryear{Alcock et al.}{1997b}]{1997ApJ...491..436A} 
Alcock C., et al., 1997b, ApJ, 491, 436 
\bibitem[\protect\citeauthoryear{Alcock et al.}{2000}]{2000ApJ...542..281A} 
Alcock C., et al., 2000, ApJ, 542, 281 
\bibitem[\protect\citeauthoryear{Alcock et al.}{2001}]{2001Natur.414..617A} 
Alcock C., et al., 2001, Natur, 414, 617 
\bibitem[\protect\citeauthoryear{Allen, Peterson, 
\& Shao}{1997}]{1997SPIE.2871..504A} Allen R.~J., Peterson D.~M., Shao M., 1997, SPIE, 2871, 504 
\bibitem[\protect\citeauthoryear{Batista et 
al.}{2009}]{2009A&A...508..467B} Batista V., et al., 2009, A\&A, 508, 467 
\bibitem[\protect\citeauthoryear{Belokurov 
\& Evans}{2002}]{2002MNRAS.331..649B} Belokurov V.~A., Evans N.~W., 2002, MNRAS, 331, 649 
\bibitem[\protect\citeauthoryear{Boden, Shao, 
\& van Buren}{1998}]{1998ApJ...502..538B} Boden A.~F., Shao M., van Buren D., 1998, ApJ, 502, 538 
\bibitem[\protect\citeauthoryear{Cassan et 
al.}{2006}]{2006A&A...460..277C} Cassan A., et al., 2006, A\&A, 460, 277 
\bibitem[\protect\citeauthoryear{Delplancke}{2008}]{2008NewAR..52..199D} 
Delplancke F., 2008, NewAR, 52, 199 
\bibitem[\protect\citeauthoryear{Dominik 
\& Sahu}{2000}]{2000ApJ...534..213D} Dominik M., Sahu K.~C., 2000, ApJ, 534, 213 
\bibitem[\protect\citeauthoryear{Fouqu{\'e} et 
al.}{2010}]{2010A&A...518A..51F} Fouqu{\'e} P., et al., 2010, A\&A, 518, A51 
\bibitem[\protect\citeauthoryear{Gould}{1992}]{1992ApJ...392..442G} Gould 
A., 1992, ApJ, 392, 442 
\bibitem[\protect\citeauthoryear{Gould}{1994}]{1994ApJ...421L..71G} Gould 
A., 1994, ApJ, 421, L71 
\bibitem[\protect\citeauthoryear{Gould}{2000}]{2000ApJ...542..785G} Gould 
A., 2000, ApJ, 542, 785 
\bibitem[\protect\citeauthoryear{Goullioud et 
al.}{2008}]{2008SPIE.7013E.151G} Goullioud R., Catanzarite J.~H., Dekens 
F.~G., Shao M., Marr J.~C., IV, 2008, SPIE, 7013,  
\bibitem[\protect\citeauthoryear{Hog, Novikov, 
\& Polnarev}{1995}]{1995A&A...294..287H} Hog E., Novikov I.~D., Polnarev A.~G., 1995, A\&A, 294, 287 
\bibitem[\protect\citeauthoryear{Hosokawa et 
al.}{1993}]{1993A&A...278L..27H} Hosokawa M., Ohnishi K., Fukushima T., Takeuti M., 1993, A\&A, 278, L27 
\bibitem[\protect\citeauthoryear{Jeong, Han, 
\& Park}{1999}]{1999ApJ...511..569J} Jeong Y., Han C., Park S.-H., 1999, ApJ, 511, 569 
\bibitem[\protect\citeauthoryear{Jiang et al.}{2004}]{2004ApJ...617.1307J} 
Jiang G., et al., 2004, ApJ, 617, 1307 
\bibitem[\protect\citeauthoryear{Koz{\l}owski et 
al.}{2007}]{2007ApJ...671..420K} Koz{\l}owski S., Wo{\'z}niak P.~R., Mao 
S., Wood A., 2007, ApJ, 671, 420 
Alcock C., Werner M.~W., Fazio G.~G., 2006, ApJ, 652, L97 
\bibitem[\protect\citeauthoryear{Lee et al.}{2009}]{2009ApJ...695..200L} 
Lee C.-H., Riffeser A., Seitz S., Bender R., 2009, ApJ, 695, 200 
\bibitem[\protect\citeauthoryear{Lindegren et 
al.}{1994}]{1994SPIE.2200..599L} Lindegren L., et al., 1994, SPIE, 2200, 
599 
\bibitem[\protect\citeauthoryear{Mao 
\& Witt}{1998}]{1998MNRAS.300.1041M} Mao S., Witt H.~J., 1998, MNRAS, 300, 1041 
\bibitem[\protect\citeauthoryear{Miyamoto 
\& Yoshii}{1995}]{1995AJ....110.1427M} Miyamoto M., Yoshii Y., 1995, AJ, 110, 1427 
\bibitem[\protect\citeauthoryear{Quirrenbach et 
al.}{1998}]{1998SPIE.3350..807Q} Quirrenbach A., et al., 1998, SPIE, 3350, 
807 
\bibitem[\protect\citeauthoryear{Takahashi}{2003}]{2003ApJ...595..418T} 
Takahashi R., 2003, ApJ, 595, 418 
\bibitem[\protect\citeauthoryear{Tisserand et 
al.}{2007}]{2007A&A...469..387T} Tisserand P., et al., 2007, A\&A, 469, 387 
\bibitem[\protect\citeauthoryear{Walker}{1995}]{1995ApJ...453...37W} Walker 
M.~A., 1995, ApJ, 453, 37 
\bibitem[\protect\citeauthoryear{Wyrzykowski et 
al.}{2009}]{2009MNRAS.397.1228W} Wyrzykowski {\L}., et al., 2009, MNRAS, 
397, 1228 
\bibitem[\protect\citeauthoryear{Yee et al.}{2009}]{2009ApJ...703.2082Y} 
Yee J.~C., et al., 2009, ApJ, 703, 2082 
\bibitem[\protect\citeauthoryear{Yoo et al.}{2004}]{2004ApJ...603..139Y} 
Yoo J., et al., 2004, ApJ, 603, 139 
\bibitem[\protect\citeauthoryear{Zub et al.}{2009}]{2009arXiv0912.2312Z} 
Zub M., et al., 2009, arXiv, arXiv:0912.2312 
\end{thebibliography}
\end{document}